\documentclass[a4paper, amsfonts, amssymb, amsmath, reprint, showkeys, nofootinbib, twoside, superscriptaddress, longbibliography, aps, pra]{revtex4-2}
\usepackage[english]{babel}
\usepackage[left=18mm,right=18mm,top=35mm,columnsep=15pt]{geometry} 
\usepackage{graphicx}
\usepackage[utf8]{inputenc}
\usepackage[colorinlistoftodos, color=green!40, prependcaption]{todonotes}
\usepackage{amsthm}
\usepackage{mathtools}
\usepackage{physics}
\usepackage[dvipsnames]{xcolor}
\usepackage{adjustbox}
\usepackage{placeins}
\usepackage[T1]{fontenc}
\usepackage{lipsum}
\usepackage{csquotes}
\usepackage{siunitx}
\usepackage{caption}
\usepackage{subcaption}
\usepackage{comment}
\usepackage{tabularx}
\usepackage[normalem]{ulem}
\usepackage{hyperref}
\usepackage[capitalize,nameinlink]{cleveref}
\crefformat{equation}{#2\textcolor{blue}{Eq.~(#1)}#3}
\Crefformat{equation}{#2\textcolor{blue}{Equation~(#1)}#3}
\crefmultiformat{equation}
  {#2\textcolor{blue}{Eqs.~(#1)}#3}  
  { and #2\textcolor{blue}{(#1)}#3} 
  {, #2\textcolor{blue}{(#1)}#3}    
  {, and #2\textcolor{blue}{(#1)}#3}
\Crefmultiformat{equation}
  {#2\textcolor{blue}{Equations~(#1)}#3}
  { and #2\textcolor{blue}{(#1)}#3}
  {, #2\textcolor{blue}{(#1)}#3}
  {, and #2\textcolor{blue}{(#1)}#3}
\usepackage{multirow}
\hypersetup{colorlinks}

\newcommand{\np}{_\text{NP}}
\newcommand{\co}{_\text{CC}}

\renewcommand{\a}{\hat{a}}
\newcommand{\ad}{\hat{a}^\dagger}
\renewcommand{\b}{\hat{b}}
\newcommand{\bd}{\hat{b}^\dagger}
\newcommand{\azi}{\langle z_\text{i} \rangle}
\newcommand{\azn}{\langle z_\text{NP} \rangle}
\newcommand{\zn}{z_{0,\text{NP}}}
\newcommand{\zi}{z_{0,\text{i}}}

\begin{document}
\title{Proposal for creating spatial superposition of a large mass in a RF trap}

\author{Martine Schut}
    \affiliation{Centre for Quantum Technologies, National University of Singapore, 3 Science Drive 2, 117543, Singapore}

\author{Jackson Tiong}
    \affiliation{Centre for Quantum Technologies, National University of Singapore, 3 Science Drive 2, 117543, Singapore}

\author{Valerio Scarani}
    \affiliation{Centre for Quantum Technologies, National University of Singapore, 3 Science Drive 2, 117543, Singapore}
    \affiliation{Department of Physics, National University of Singapore, 2 Science Drive 3, 117542, Singapore}


\begin{abstract}
Engineering coherent spatial superpositions of levitated large masses is an ongoing challenge. Borrowing from recent experimental work, we consider a charged mass of hundreds of nanometers size (``nanoparticle'') co-trapped with an ion in a Paul trap, and propose a scheme to manipulate its spatial state through the Coulomb interaction with the ion. We focus on the achievable delocalisation, only sketching the other challenges of the protocol (initial cooling, preservation of coherence for long-enough times, and detection). We prove that our scheme can displace coherently the nanoparticle by a few nanometers, and give an estimate of the allowed noise. Though smaller than the nanoparticle's size, the nanoparticle displacement is much larger than the wavefunction of the trap's ground state. Thus the co-trapping scheme is in principle able to demonstrate the delocalisation of a charged nanoparticle if the required noise isolation can be realised.
\end{abstract}


\maketitle

\section{Introduction}\label{sec:intro}
One of the core principles of quantum physics is that of superposition of states. 
This principle is widely used in quantum technologies and quantum information, but is not observed in our everyday macroscopic world.
By creating quantum superpositions in increasingly large objects, we are pushing the boundary between quantum and classical physics.
We focus on the quantum superposition of two states localized at different positions, i.e. spatial delocalisation.
These spatial superpositions have been realised in small particles and have found applications in, for example, the Colella-Overhauser-Werner (COW) experiment for showing that the equivalence principle holds in the quantum limit~\cite{colella_observation_1975}, the challenge now is to have both a large mass and a large superposition size.
Increasing the mass is particularly important for applications related to gravity, which scales with mass.
A large spatial superposition ($0.54$~m) of a very small particle~\cite{kovachy2015quantum}, and a small cat states in heavy masses ($16~\mu$g)~\cite{bild_schrodinger_2023}, have been reported; but the regime of heavy masses with large superpositions (at least attograms and micro/nanometers) has not yet been reached experimentally.
These macroscopic quantum systems have proposed applications in testing fundamental physics: they explore the boundary between classical and quantum physics and can test foundations of quantum mechanics (e.g., CSL and Di\'{o}si-Penrose models~\cite{Bassi_2013,howl2019exploring}). 
They could also have multiple applications in sensing, such as for small force measurements such as from the gravitational interaction~\cite{bose2017spin,marletto2017gravitationally} or dark matter interaction~\cite{Riedel_2013}; and have been suggested for applications in geophysics, volcanology and seismology, since they can be used for sensing density shifts~\cite{stray2022quantum}.

We are particularly interested in levitated systems, since they are less connected to the environment and have potential for detecting superpositions of the gravitational field.
There are several platforms that work towards realising macroscopic quantum superpositions, such as in optically trapped nanoparticles, diamagnetically levitated systems and free-fall experiments.
The first one has proven successful for atomic and nanoparticles in terms of cooling~\cite{delic2020cooling,gonzalez2021levitodynamics} but suffers a lot of decoherence due to heating from the trapping laser field.
In diamagnetically levitated spin-based platforms~\cite{margalit2021realization,Bose:2025qns}, good spin manipulation and readout have been achieved~\cite{Walsworth2020}.
Protocols for creating superpositions in these different platforms can vary as well. 
Near-field interferometry has recently improved their macroscopicity, see Ref.~\cite{pedalino2025probing}, with promises of increasing the macroscopicity further.

In this paper, we focus on electromagnetically trapped charged systems, where a large electric dipole moment or a large number of charges on the test mass can be used to cool the system in three degrees of freedom~\cite{Dania2021cooling,Barker:2023,martinetz2021electric}, and near the motional ground state for optically trapped nanoparticles~\cite{delic2020cooling,troyer2025quantum}. 
Creating large coherent cat states has been an active research area in the ion trapping community since the '90s, both in single or multiple Paul/Penning traps~\cite{monroe_schrodinger_1996,schmidt2003realization,johnson_ultrafast_2017}.
The internal structure of the atomic particle can be manipulated with a series of laser pulses. After preparing a superposition of internal states, these laser pulses can be used to give the atomic particle a state-dependent kick and thus create a spatial superposition.
As particles increase in size, their internal structure becomes more complicated and, consequently, the direct application of these methods to large masses, i.e. dielectric media, becomes challenging~\cite{arrangoiz2019resolving}.
Inspired by recent work~\cite{bykov2024nanoparticle}, our proposal is based on a scheme where an ion and the nanoparticle are co-trapped in a dual frequency Paul trap.
One can first prepare a spatial superposition of the ion with standard methods, then let it induce a correlated displacement of the nanoparticle via their mutual distance-dependent interaction. This approach avoids the problem of accessing the nanoparticle internal energy structure, while still having the cooling advantage of a charged nanoparticle.
This methodology is similar to the phonon-induced displacements in nanomechanical resonators such as in ref.~\cite{Pathak2017}, the hybrid optical-electromagnetic trap for an atom-nanoparticle systems such as in ref.~\cite{torovs2021creating}, and the nanoparticle coupled to a superconducting qubit in a different geometry Paul trap such as in ref.~\cite{martinetz2020quantum}.
Like the latter setup, this work avoids the dominant decoherence channel sourced by optical fields.
Compared to schemes using grating interferometry~\cite{pedalino2025probing,bateman2014near}, this setup is not limited by a set interaction (e.g. Talbot) time.

In this work we show that there exists a parameter regime, in which sizeable spatial superpositions can in principle be prepared. We also give an initial analysis of the challenges related to the readout of the superposition and coherence loss. 
We start by introducing our setup and give a brief background on linear Paul traps driven by two frequencies for trapping the nanoparticle-atomic ion system (~\cref{sec:setup}). We present the proposed protocol as well as exemplary parameters that could be used for experimental realisation. 
\Cref{sec:quantum} then discusses the coherent delocalisation and the resulting induced nanoparticle superposition width as well as a measurement scheme. 
Finally, these results are repeated taking into account noise sources such as linear electric field noise, vibrational noise, improper closure of the interferometer, non-ground-state initial states and air molecule scattering (~\cref{sec:noise}).

\section{Platform: co-trapping setup}\label{sec:setup}

\begin{figure}
    \centering
    \includegraphics[width=0.85\linewidth]{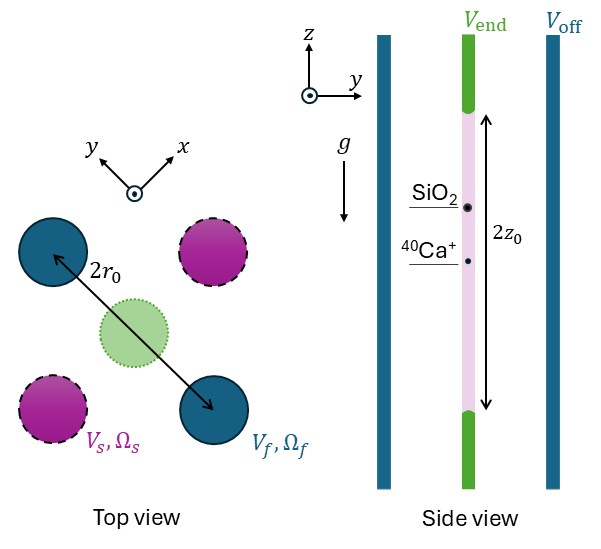}
    \caption{Exemplary schematic of the setup, showing a linear dual frequency Paul trap. The blue electrodes are driven by the fast voltage while the purple-colour electrodes are driven by the slow voltage. The green-colour electrodes are the end-cap electrodes driven by a DC voltage. The distances and co-trapped particles are indicated in the figure; the figure is not to scale. Such a setup was proposed an realized in Ref.~\cite{bykov2024nanoparticle}.}
    \label{fig:Paul-trap}
\end{figure}

We consider the co-trapping of a charged nanoparticle and an atomic ion in a dual frequency linear Paul (RF) trap; see the schematic in \cref{fig:Paul-trap}. This is obtained by driving the Paul trap with two oscillating electric fields, one with slow frequency ($\Omega_s$) that provides the trapping for the nanoparticle, and one with high frequency ($\Omega_f$) that provides the trapping of the ion. 
After the introduction of the Paul trap~\cite{Paul:1990}, dual frequency traps have been proposed as a method for co-trapping~\cite{Trypogeorgos2016} of particles with very different charge-to-mass ratios.
Demonstrated in ref.~\cite{bykov2024nanoparticle} for an atomic ion and a charged silica nanoparticle, this co-trapping is in principle stable for a range of experimental parameters, which can either be found numerically or approximated analytically. For details on the latter we refer to~\cref{app:stability}.

The radial, time-dependent electric potential of the trapping electric field is given by
\begin{align}
    V_\text{rad}(x,y,t) = &- \left( V_s \cos(\Omega_s t) + V_f \cos(\Omega_f t) \right) \frac{x^2-y^2}{2 r_0^2} \nonumber \\
    & - V_\text{off} \, \frac{x^2 - y^2}{2 r_0^2} - V_\text{end} \, \frac{x^2 + y^2}{2 z_0^2} \label{eq:pot-radial}
\end{align}
with $V_s$ and $V_f$ the slow and fast RF voltages respectively, $V_\text{off}$ an offset voltage from the radial electrodes, where $2 r_0$ the distance between the radial electrodes, and $V_\text{end}$ the end-cap voltage~\footnote{To account for the non-ideal shape of the electrodes, the voltages can be scaled by a factor $\kappa$; we have assumed hyperbolic end-caps, i.e. $\kappa=1$ here, but later on imperfections in the RF electrodes are accounted for by scaling with $\kappa_\text{RF} = 0.93$ and $\kappa_\text{end}=0.22$ for the end-caps.\label{footnote:kappa}}. 
The RF field is assumed zero along the z-axis, allowing for trapping along this axis: the confinement in the z-direction is provided by the static end-cap voltages $V_\text{end}$.
\begin{equation}\label{eq:pot-axial}
    V_\text{ax}(z) = V_\text{end} \, \frac{z^2}{z_0^2} 
\end{equation}
Here, $2z_0$ is distance between the axial electrodes.
In the ideal case where the rods have perfect quadrupole symmetry and are long compared to their separation, the offset voltage is constant along the z-axis close to the trap centre.
Combining~\cref{eq:pot-radial,eq:pot-axial}, the Laplace equation is satisfied, which means that the potential will be harmonic and the particle(s) can be trapped stably, with dynamical confinement in the $xy$-plane and static confinement in the $z$-direction~\cite{Paul:1990}.

For a particle of mass $m$ and charge $Q$ (either the nanoparticle or the ion), the equations of motion read
\begin{align}
    m \ddot{u} &= - Q \pdv{V_\text{rad}}{u} - \pdv{U_\text{I}}{u}\qq{} u = x, y \label{eq:eom-x-gen}\\
    m \ddot{z} &= - Q \pdv{V_\text{ax}}{z} - \pdv{U_\text{I}}{z} -  \pdv{U_\text{g}}{z} \label{eq:eom-z-gen}
\end{align}
where $V_\text{rad}$ and $V_\text{ax}$ are the radial and axial electric potentials given in~\cref{eq:pot-radial,eq:pot-axial}; 
$U_\text{g}$ is the gravitational potential energy, assuming that the $z$ axis of the trap is aligned with the local gravitational field for simplicity; and $U_\text{I}$ is the interaction potential energy between the masses, discussed below.

The interaction potential $U_\text{I}$ includes several interactions, but it will be dominated in this case by the Coulomb interaction between the two charges (see~\cref{app:interactions}).
We will expand the Coulomb interaction $U\co = \frac{1}{4\pi\epsilon_0}\frac{Q_\text{i} Q\np}{d}$ in the limit where the oscillations are small compared to the equilibrium separation $z\np^\text{eq} - z_\text{i}^\text{eq}$ and assume (see \cref{sec:setup}) that the two trapped masses have the same equilibrium position for the $x$- and $y$-coordinates, and so their separation distance is given by:
\begin{align}
    d(t)^2 &= (z\np^\text{eq} - z_\text{i}^\text{eq} + \delta z\np(t) - \delta z_\text{i}(t))^2 \label{eq:distance} \\ 
    &+ (\delta x\np(t) - \delta x_\text{i}(t))^2 + (\delta y\np(t) - \delta y_\text{i}(t))^2\,. \nonumber 
\end{align} 
The linearized Coulomb interaction is:
\begin{widetext}
\begin{align}
    U_\text{I}\,\approx\, U\co \,\approx
    &\frac{Q_\text{i} Q\np}{4\pi\epsilon_0}\frac{1}{z^{\text{eq}}\np - z_\text{i}^\text{eq}} \bigg[ 1 - \frac{\delta z\np - \delta z_\text{i}}{z^{\text{eq}}\np - z_\text{i}^\text{eq}} + \frac{2(\delta z\np - \delta z_\text{i})^2 - (\delta x\np - \delta x_\text{i})^2 - (\delta y\np - \delta y_\text{i})^2}{2(z^{\text{eq}}\np - z_\text{i}^\text{eq})^2} + \ldots \bigg] \label{eq:lin-coulomb}
\end{align}
\end{widetext}


The dual RF trap setup as described has been realised experimentally in ref.~\cite{bykov2024nanoparticle}, which was the inspiration for this work.
Specifically, they demonstrated the the confinement of a silica nanoparticle and an atomic calcium ion in the same dual frequency trap, where trapping of the nanoparticle is dominated by the slow field and the stable trapping of the ion imposes constraints on the fast and slow fields.
Although the trapping is initially done in the $xy$-plane, by adjusting the voltages on the off-set and end-cap electrodes, the alignment of the co-trapped pair can be rotated to have separation solely in the $z$-direction, with the ion at the origin.
The excess micro-motion~\cite{berkeland1998minimization} and sensitivity to voltage noise is better for this latter configuration~\cite{bykov2024nanoparticle} since the end-caps have a DC voltage. It is therefore assumed to be the setup for this work.
This configuration also eliminates the interaction in the radial plane, since there is no (or negligible) separation in the radial direction.

\Cref{table:params} gives an overview of the trapping parameters, which are mostly based on the experimental realisation of ref.~\cite{bykov2024nanoparticle}, and the resulting separation and frequencies.

\begin{table}[bpt]
\centering
\setlength{\tabcolsep}{10pt}
\def\arraystretch{1.5}
  \begin{tabular}{l l}
 Setup params & value \\ [0.5ex]
 \hline 
 \hline
 $\Omega_s$, $\Omega_f$ & $7$~kHz, $17.5$~MHz \\ 
 \hline 
 $V_s$, $V_f$ & $80$~V, $1250$~V \\ 
 \hline 
 $r_0$, $z_0$ & $0.9$~mm, $3.4$~mm \\ 
 \hline
 $\kappa_\text{RF}$, $\kappa_\text{end}$~\ref{footnote:kappa} & $0.93$, $0.22$ \\
 \hline
 $V_\text{end}$ & $200-400$~V \\
 \hline 
 $Q\np$, $Q_\text{i}$  & $800e$, $e$ \\ 
 \hline 
 $m\np$  & $5\times10^{-16}$~kg  \\ 
 \hline
 \hline
 & \\ [-0.5ex]
 Resulting params & value \\ [0.5ex]
 \hline 
 \hline
 Separation NP-ion & $55-43~\mu$m \\
 \hline
 $\omega\np \, (x, y, z)$ & $(1, 1, 2)$~kHz \\
 \hline 
 $\omega_\text{i} \, (x, y, z)$  & $(2, 4, 1)$~MHz \\
 \hline 
  \end{tabular}
  \caption{Overview of proposed experimental parameters, based on the dual frequency Paul trap in ref.~\cite{bykov2024nanoparticle}, and the parameters resulting from this setup.
  We consider a difference in charge-to-mass ratio of $\sim10^6$ and a difference in mass ratio of $\sim10^8$, where the nanoparticle mass is taken to be $5\times10^{-16}\,\si{\kilogram}$.
  The separation between the ion (i) and the nanoparticle (NP) depends on the end-cap voltage and is given for the non-interaction case $55~\mu$m ($200$~V) to the smallest separation considered here: $43~\mu$m ($400$~V).
  The frequencies are calculated as explained in~\cref{app:stability}, which  also gives the stability conditions of the trapping, which are satisfied for the values in this table.}
  \label{table:params}
\end{table}

The observations of ref.~\cite{bykov2024nanoparticle} demonstrate that the mutual interaction is sufficient to induce displacements for a distance  $z\np^\text{eq} - z_\text{i}^\text{eq}\lesssim 50~\si{\micro\metre}$, beyond which trapping dominates. We propose to utilise this mutual interaction to induce a superposition in the nanoparticle.
Here, we only consider the dominant Coulomb interaction for the induced nanoparticle displacement. A comparison of the estimated magnitude for different electromagnetic forces, showing that the Coulomb interaction is the dominant one, is given in~\cref{app:interactions}.

\section{Coherent delocalisaton of the nanoparticle: ideal case}\label{sec:quantum}

Our goal is to propose a protocol for the coherent delocalisation of the nanoparticle the dual-frequency setup under study, exploiting the distance-dependent mutual interaction with the ion of \cref{eq:lin-coulomb}. 

\subsection{Overview of the protocol}\label{subsec:steps}

We sketch the protocol here, followed by the equations of the desired operations in the ideal case, when both the nanoparticle and the ion are initialised in the motional ground state and there is no decoherence, which is the blueprint for the idea. Thermal excitations and decoherence mechanisms will be considered in~\cref{sec:noise}.

\textit{Step 1: Loading and cooling the two objects.} This consists of three operations. First, loading and cooling the nanoparticle. In the protocol described in ref.~\cite{bykov2024nanoparticle}, a nanoparticle can be loaded into the trap in a ultra-high vacuum, and motional cooling can be done with optical detection and electric feedback. Eventually there is the hope that this platform will be able to cool the system to the motional ground state. Second, loading the ion and cooling. With the off-set voltage, the nanoparticle can be moved away from the centre such that the ion can be loaded and cooled there. At this point, in the ideal case we have prepared the state 
    \begin{align}
        \ket{\psi_0} &= \ket{0}\np \ket{0}_\text{ion}\ket{\downarrow}\,. \label{eq:psi3}
    \end{align}
With a $\pi/2$ pulse on the ion, we create a superposition of two selected pseudo-spin states: 
\begin{align}
\ket{\psi_{1}} &= \hat{U}_{\pi/2}\ket{\psi_0} = \ket{0}\np \ket{0}_\text{ion}(\ket{\downarrow}+\ket{\uparrow})/\sqrt{2}\,. \label{eq:psi4a}
\end{align}

    We assume the initial separation between the two particles is $> 50\,\si{\micro\metre}$ such that their mutual interaction is negligible~\cite{bykov2024nanoparticle}. Finally, the DC voltages are adjusted such that the separation between the ion and nanoparticle is solely in the $z$-direction. Then, by adjusting the end-cap voltages, the separation between the particles is reduced such that their mutual interaction increases.

\textit{Step 2: Create the ion cat state}. 
With a spin-dependent kick (SDK) the motional states of the ion become entangled with its internal spin states, creating a cat-state\begin{align}
\ket{\psi_{2}} &=\hat{U}_\text{SDK}\ket{\psi_{1}} = \ket{0}\np(\ket{-\beta}_\text{ion}\ket{\downarrow}+\ket{\beta}_\text{ion}\ket{\uparrow})/\sqrt{2}\,, \label{eq:psi4b}
\end{align}
where $\beta$ is the spin-dependent displacement of the motional state of the ion.

\textit{Step 3: Inducing delocalisation of the nanoparticle}. The SDK shifts the equilibrium position of the ion in a spin-dependent way. Through the Coulomb interaction, this shifts the equilibrium position of the co-trapped nanoparticle. Assuming that these changes are instantaneously effective, in the new natural spin-dependent modes the state reads
\begin{align}
    \ket{\psi_{2}'} &=\hat{U}_\text{C}\ket{\psi_2}
    = (\ket{-c_a^\downarrow}\np\ket{-\beta-c_b^\downarrow}_\text{ion}\ket{\downarrow} \nonumber \\ &\qq{}\qq{}\quad+\ket{-c_a^\uparrow}\np\ket{\beta-c_b^\uparrow}_\text{ion}\ket{\uparrow})/\sqrt{2} \label{eq:psi5} 
\end{align}
where $c_a^{\uparrow/\downarrow}$ is the amount of displacement felt by the nanoparticle, and $c_b^{\uparrow/\downarrow}$ is the same for the ion. Being out of equilibrium, the state evolves as $\ket{\psi'_3(t)}$, which can be understood as a coherent superposition of oscillations in either trap. In particular, in the ideal case this gives rise to a periodic separation and recombination of the two wavefunctions; we denote this period by $T$. The separation is maximal $t=T/2$ (modulo $T$). At those times,  
\begin{align}
    \ket{\psi_{3}'(T/2)} &=\hat{U}_\text{C}\ket{\psi_2}
    = (\ket{+c_a^\downarrow}\np\ket{-\beta+c_b^\downarrow}_\text{ion}\ket{\downarrow} \nonumber \\ &\qq{}\qq{}\quad+\ket{+c_a^\uparrow}\np\ket{\beta+c_b^\uparrow}_\text{ion}\ket{\uparrow})/\sqrt{2} \label{eq:psi5b} 
\end{align}

\textit{Step 4: Measurement.} In this paper, we consider an interferometric test of the coherence. A uniform electric field induces a displacement $\xi_c$ and, more importantly, a position-dependent phase on the nanoparticle (the same effects on the ion are much smaller and we neglect them). If we assume that this process is instantaneous and performed at the time when the separation is maximal, we get
\begin{align}
    \ket{\psi'_{4a}}
    &= \hat{U}_{\text{control}}\ket{\psi'_3(T/2)} \nonumber \\
    &= (\ket{c_a^\downarrow+\xi_c}\np\ket{-\beta+c_b^\downarrow}_\text{ion}\ket{\downarrow} \label{eq:psi6a} \\ &\quad+e^{i\phi(c_a^\uparrow,c_a^\downarrow)}\ket{c_a^\uparrow+\xi_c}\np\ket{\beta+c_b^\uparrow}_\text{ion}\ket{\uparrow})/\sqrt{2}. \nonumber
\end{align} Then, after another half period, with a reverse SDK the coherence is transferred back to the spin, which in the original modes leads to \begin{align}
\ket{\psi_{4b}}
    &=\hat{U}_\text{i-SDK}\hat{U}^\dagger_\text{C}\ket{\psi'_{4a}}\nonumber\\
    &= \ket{\xi_c}\np\ket{0}_\text{ion}(\ket{\downarrow}+e^{i\phi(c_a^\uparrow,c_a^\downarrow)}\ket{\uparrow})/\sqrt{2}\,. \label{eq:psi6b}
\end{align} A measurement in the complementary basis $\ket{\pm}=\frac{1}{\sqrt{2}}(\ket{\uparrow}\pm \ket{\downarrow})$ typically yields \begin{align}
    P_\pm = \frac{1}{2} \pm \frac{1}{2} V\cos(\phi) \label{eq:prob}
\end{align} with $V\approx 1$ in the ideal case. One remark: perfect spin coherence could be achieved without ever transferring any information on the other degrees of freedom; but the phase imprinted by the electric field on the ion is $10^3$ times smaller than that imprinted on the nanoparticle. Thus the interference pattern is sufficient to show that the coherence was imprinted on the nanoparticle. As an additional test, one may want use some runs to check that the positions of the ion and the nanoparticle are (classically) correlated at the end of Step 3.

Finally let us note that other detection schemes could be used, like ref.~\cite{BoseKim2011}, where the coupling to a qubit is used to reconstruct the nanoparticle characteristic function could also be applicable by co-trapping a second ion; or a Talbot-Lau-type scheme such as in ref.~\cite{nimmrichter_electron-enabled_2025}.

\subsection{Steps 1 and 2: initialisation and manipulation of the ion}\label{subsec:prep} 

On the nanoparticle side, we have nothing to add, besides restating that in this section we assume its motional state to be the ground state, a very challenging assumption that will be relaxed in \cref{sec:noise}.

The steps involving the ion are standard in ion trapping. 
The ion's motion is initialised by the Doppler cooling and subsequent resolved sideband cooling brings the average vibrational occupation number $\bar{n}_0\approx 0$~\cite{diedrich1989laser}.
The atomic ion considered here, $^{40}\text{Ca}^+$, has several electronic states available to encode the a pseudo-spin.
Typically, the qubit is encoded in the ground and metastable states:
\begin{equation}\label{eq:spin-qubit}
    \ket{\downarrow}\equiv\ket{S_{1/2},M_J=1/2}\, , \qq{}\ket{\uparrow}\equiv\ket{D_{5/2},M_J=3/2} \, .
\end{equation}
A ground state and a meta-stable state (with a lifetime of $\sim1$~s~\cite{barton2000measurement}) with a transition at $729$~nm; an applied magnetic field causes a linear split in the energy levels, lifting the Zeeman degeneracy.
The advantage of these internal states is that their transition frequency is far removed from other internal state transitions for an applied magnetic field of $119.6$~G (by $200$~MHz~\cite{Lo:2014vda,lo2015creation}), they are not as sensitive to magnetic field fluctuations and the long lifetime of the meta-stable state reduces problems with spontaneous emission.
The disadvantage is that their transitions happen via the quadrupole moment and have a very small linewidth, i.e., lasers with very stable frequencies are required~\cite{lo2015creation}.
For example, ref.~\cite{Lo:2014vda} demonstrated spin-motion entanglement in $^{40}\text{Ca}^+$ using these pseudo-spin states. In the following we discard these details and simply consider an encoded qubit state in the ion.

A spin-dependent kick (SDK) prepares the cat state \cref{eq:psi4b} of the spin and motional state of the ion~\cite{monroe_schrodinger_1996}. 
Although the SDK is a time-dependent pulse, due to the short duration of the pulse (compared to the oscillation frequencies), we consider it to be instantaneous in this work. 
Creating large cat states in single trapped ions is widely studied and complex protocols have been developed for creating fast and large cat states, such as for example in ref.~\cite{johnson_ultrafast_2017} for $^{171}\text{Yb}^+$.

The evolution operator from the SDK is a state-dependent displacement operator accompanied by a spin-flip~\cite{lee_phase_2005,wineland_experimental_1998,wineland_quantum_2003}:
\begin{equation}\label{eq:SDK}
    \hat{U}_\text{SDK} = 
    \hat{D}_\text{i}(\beta) \ket{\uparrow}\bra{\downarrow} +  \hat{D}_\text{i}(- \beta) \ket{\downarrow}\bra{\uparrow} \, ,
\end{equation}
$\hat{D}_\text{i}$ is the phase–space displacement operator on the ion.
Such protocols have been realised for different ion species and with different characteristics for the laser pulses \cite{johnson_ultrafast_2017,wineland_experimental_1998,wineland_quantum_2003,lee_phase_2005,Lo:2014vda}.
These experiments indicate that a displacement $\abs{\beta} = 0.1-10$ can be achieved, which corresponds to $\Delta z_\text{i} \approx 1-100$~nm for the width of the ground state wavefunction is given by $z_{0,i}=\sqrt{\hbar/2m\omega_z}\approx 4\,\si{\nano\metre}$ (much smaller than the wavelength of the laser fields, i.e. the Lamb-Dicke regime).
Ref.~\cite{lo2015creation}, which used the spin qubit as in \cref{eq:spin-qubit}, showed a maximum displacement of $\abs{\beta}>19$, although this protocol considered an SDK pulse of $250~\mu$s, which is larger than the oscillation period of the ion. To our knowledge, an ultrafast protocol specified to the calcium ion has yet to be shown experimentally.
Ref.~\cite{liu2025high} studied the application of SDKs in the presence of micromotion.

\subsection{Step 3: conditional displacement of the nanoparticle}\label{sec:displacement} 
\begin{figure}[t]
    \centering
    \includegraphics[width=0.6\linewidth]{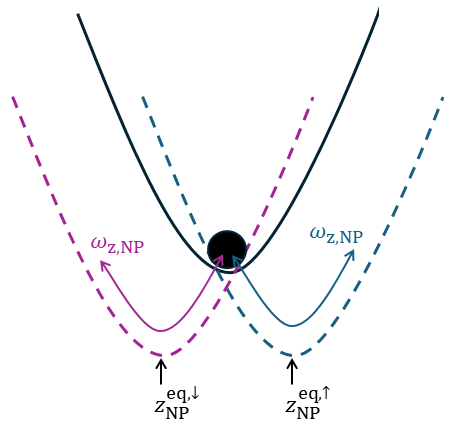}
    \caption{The Coulomb interaction induces a superposition of equilibrium positions $z^{\text{eq},\uparrow,\downarrow}\np$ (purple/blue) in the initial state (black) of the nanoparticle, due to the ion spin-dependent superposition.}
    \label{fig:physics}
\end{figure}

Considering the SDK to be effectively instantaneous, the entangling ion-nanoparticle interaction can be thought of as being "turned on". A pictorial understanding of the situation is given in \cref{fig:physics}. The nanoparticle, that was at rest before the SDK was applied, finds itself out of equilibrium in one of the two possible traps, depending on the spin state, and starts an oscillatory motion. This physical understanding is similar to the spin-based platform protocols for creating spatial superposition such as in ref.~\cite{pedernales_motional_2020}

The effective Hamiltonian that governs the interaction can be written
\begin{equation}\label{eq:H-spin-dep}
    \hat{H}_\text{T} = \hat{H}(d_\text{eq}^\uparrow) \ket{\uparrow}\bra{\uparrow} + \hat{H}(d_\text{eq}^\downarrow) \ket{\downarrow}\bra{\downarrow}
\end{equation}
where $\hat{H}(d_\text{eq})$ indicates the Hamiltonian arising from the Coulomb interaction \cref{eq:lin-coulomb} for an equilibrium distance $d_\text{eq}$ given by:
\begin{align}
    d_\text{eq}^\uparrow &= \azn - \azi + 2 \zi \Re{\beta} \\
    d_\text{eq}^\downarrow &= \azn - \azi - 2 \zi \Re{\beta}
\end{align}
where $\zi = \sqrt{\hbar/(2m_\text{i}\omega_{z,\text{i}})}$ is the zero-point width for the ion and $\langle z \rangle$ is the equilibrium position before the SDK and the effective Coulomb interaction.

As usual, we quantize the interaction by quantizing the fluctuations around the equilibrium position for the ion $\delta\hat{z}_\text{i} = z_{0,i}(\hat{b}^\dagger+\hat{b})$ and for the nanoparticle $\delta\hat{z}\np = z_{0,\text{NP}}(\hat{a}^\dagger+\hat{a})$. Inserting these expressions into \cref{eq:lin-coulomb}, and neglecting the terms of order $O(z_0^2/d_\text{eq}^3)$ (justified in \cref{app:coulomb}), we find
\begin{align}
    \hat{H}(d_\text{eq}) =& \hat{H}\np(d_\text{eq})+\hat{H}_\text{i}(d_\text{eq})
\end{align}
with  
\begin{align} 
&\hat{H}_k(d_\text{eq})= - \frac{\hbar \omega_k}{4} \left(\hat{a}_k^\dagger - \hat{a}_k\right)^2 +
    Q_k \frac{V_\text{end}}{z_0^2}z_k^2 \left( \a_k + \ad_k \right)^2 \nonumber \\
    &+\left[ 2 Q_k \frac{V_\text{end}}{z_0^2} z^\text{eq}_k - \frac{1}{4\pi\epsilon_0}\frac{Q_\text{i} Q\np}{d_\text{eq}^2} + m g \right] z_{0,k} \left( \hat{a}_k + \hat{a}^\dagger_k\right) \label{eq:hd}
\end{align}
the first terms being the kinetic energy, for either object $k=\textrm{NP},\textrm{i}$ (and the operators $\hat{a}_{\textrm{NP}}\equiv \hat{a}$ and $\hat{a}_{\textrm{i}}\equiv \hat{b}$). Notice that each of the terms in \cref{eq:H-spin-dep} is a non-interacting Hamiltonian between the two modes: the three-body entanglement \cref{eq:psi3} will be created by the spin-dependence, as illustrated in \cref{fig:physics}.

Now we rewrite this Hamiltonian. First, we perform a spin-dependent redefinition of the modes: $\hat{a}_{\uparrow/\downarrow} = \hat{a} + c_a^{\uparrow/\downarrow}$, $\hat{b}_{\uparrow/\downarrow} = \hat{b} + c_b^{\uparrow/\downarrow}$ (see~\cref{app:calc-shift} for the expressions $c_a$, $c_b$; which may be time-dependent is the SDK were not instantaneous).
Effectively, this is a displacement operation since $D^\dagger(c_a) a D(c_a) = a + c_a$; so this change of modes is equivalent to
\begin{align}\label{eq:coulomb SDD}
    \hat{U}_\text{C} &= 
    \hat{D}\np(-c_a^\uparrow)\otimes \hat{D}_\text{i}(-c_b^\uparrow)\ket{\uparrow}\bra{\uparrow}\\ &\quad+  \nonumber \hat{D}\np(-c_a^\downarrow)\otimes\hat{D}_\text{i}(-c_b^\downarrow) \ket{\downarrow}\bra{\downarrow} \, ,
\end{align}
and the state becomes \cref{eq:psi5} in the new spin-dependent modes. Second, we diagonalize the Hamiltonian via a Bogoliubov transformation (see~\cref{app:BT} for more details) such that \cref{eq:H-spin-dep} becomes
\begin{align}
    \hat{H}_T = \tilde{C} &+ \underbrace{\hbar\left(\tilde{\omega}_a \tilde{a}_\uparrow^\dagger\tilde{a}_\uparrow + \tilde{\omega}_b \tilde{b}_\uparrow^\dagger\tilde{b}_\uparrow\right)}_{\tilde{H}(d_\text{eq}^\uparrow)}\ket{\uparrow}\bra{\uparrow} \nonumber\\ 
    &+ \underbrace{\hbar\left(\tilde{\omega}_a \tilde{a}_\downarrow^\dagger\tilde{a}_\downarrow + \tilde{\omega}_b \tilde{b}_\downarrow^\dagger\tilde{b}_\downarrow\right)}_{\tilde{H}(d_\text{eq}^\downarrow)}\ket{\downarrow}\bra{\downarrow}  \, ,\label{eq:ham-rewritten}
\end{align}
where $\tilde{a}$, $\tilde{b}$ are the transformed ladder operators with $[\tilde{a},\tilde{a}^\dagger] = 1$ and similarly for $\tilde{b}$. The frequencies of the new modes are given by
\begin{align*}
    \tilde{\omega}_a &= 2 \zn \sqrt{ Q\np \omega\np \frac{V_\text{end}}{\hbar z_0^2}} = \omega_\text{NP}\,,\\
    \tilde{\omega}_b &= 2 \zi \sqrt{Q_\text{i} \omega_\text{i} \frac{V_\text{end}}{\hbar z_0^2} } =  \omega_\text{i}
\end{align*} and so happen to coincide exactly with the frequencies of the ion and the nanopartilce. This is due to having neglected terms in the Hamiltonian. In~\cref{app:supp}, this analysis is repeated without that approximation, in which case the frequencies acquire a dependence on the displaced equilibrium positions.

\subsection{Resulting displacement of the nanoparticle}\label{sec:results}
In the Heisenberg picture, the ladder operators $\tilde{a}_{\uparrow,\downarrow}$ evolve according to the Hamiltonian in \cref{eq:ham-rewritten}:
\begin{align}
    \tilde{a}_\uparrow(t) 
    &= e^{i \tilde{H}(d_\text{eq}^\uparrow) t/\hbar} \tilde{a}_\uparrow e^{-i\tilde{H}(d_\text{eq}^\uparrow) t/\hbar} = \tilde{a}_\uparrow(0) e^{-it\tilde{\omega}_a^\uparrow/\hbar}\,. \label{eq:at}
\end{align}
and similarly for $\tilde{a}_\downarrow(t)$.
To find the evolution of the position $\hat{z}\np = \zn (\hat{a} + \hat{a}^\dagger)$, we evolve it via the Hamiltonian $\hat{H}(d_\text{eq}^{\uparrow,\downarrow})$, perform the shift and Bogoliubov transformation so that we can use the time evolution in terms of the modes $\tilde{a}$ in \cref{eq:at}, and switch back to find the evolution in terms of $\hat{a}$, $\hat{a}^\dagger$.
The displacement induced in the average position of the nanoparticle is given by the difference between the evolution via $\hat{H}(d_\text{eq}^\uparrow)$ and that via $\hat{H}(d_\text{eq}^\downarrow)$ (determined by the distance between the $\pm\beta$ ion states):
\begin{align}
    \Delta z\np(t) 
    &= e^{i \hat{H}_+ t/\hbar} \hat{z} \np e^{-i\hat{H}_+ t/\hbar} - e^{i \hat{H}_- t/\hbar} \hat{z}\np e^{-i\hat{H}_- t/\hbar}\nonumber
    \\
    & \approx 2 \zn \abs{c_a^\uparrow - c_a^\downarrow} \left[ 1 - \cos(\omega\np t) \right] \label{eq:delz-res}
\end{align}
The cosine-behaviour of the superposition size originates from the oscillatory behaviour around different equilibrium positions, as expected from \cref{fig:physics}. 

\begin{figure}[t]
    \centering
    \includegraphics[width=\linewidth]{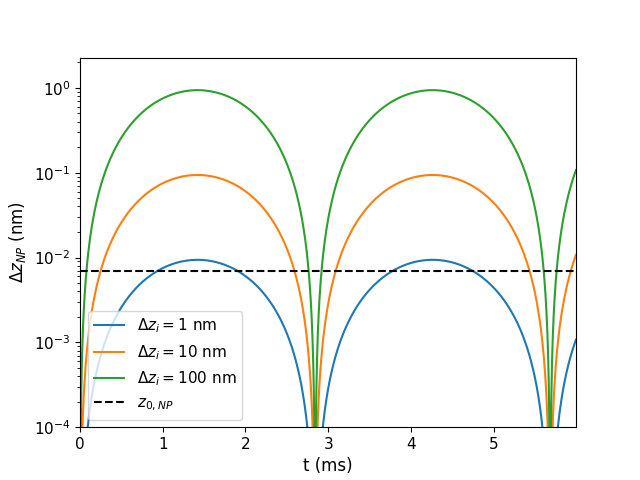}
    \caption{The induced superposition width in the nanoparticle as a function of time for different displacements in the ion equilibrium state. The end-cap voltage is taken to be $500$~V, which gives an equilibrium separation of $\sim 40\,\si{\micro\metre}$ between the ion and nanoparticle. The dashed line indicates the width of the nanoparticle's wavefunction cooled in the ground state of the trap, $z_{0,\text{NP}} = \sqrt{\hbar/(2m\np\omega_{z,\text{NP}})}$.}
    \label{fig:res}
\end{figure}

The separation is maximum at $t=\pi/\omega\np$, giving $\Delta z\np = 4 \zn \abs{c_a^\uparrow - c_a^\downarrow}$. This expression is plotted in \cref{fig:res}.
Note that the the frequency $\tilde{\omega}_a = \omega_\text{z,NP}$ in our approximation, which depends on the end-cap voltage $V_\text{end}$, charge-to-mass ratio and trap geometry ($z_0$ in \cref{fig:Paul-trap}).
The displacement values $c_a^{\uparrow,\downarrow}$ depends on the end-cap voltage and the mutual Coulomb interaction between the ion and the nanoparticle.

\Cref{eq:delz-res} is plotted in \cref{fig:res} for different displacements of the ion equilibrium position, $\Delta z_\text{i}$ (from the SDK).
This displacement is, as explained in~\cref{subsec:prep}, assumed to be approximately $1-100$~nm. To find the separation between the equilibrium positions of the ion and nanoparticle, we solve the equations of motion (\cref{eq:eom-z-gen}) numerically, given the end-cap voltage of $500$~V, which gives a separation of approximately $40\,\si{\micro\metre}$.
The other experimental parameters are as described in~\cref{table:params}.
Increasing the off-set voltage will decrease the separation due to the steeper trapping potential, which in turn will decrease the mutual separation and thus cause a stronger Coulomb coupling. Despite the stronger coupling increasing the off-set voltage will not necessarily increase the induced superposition size as the potential becomes steeper.

Equation~\cref{eq:delz-res} and \cref{fig:res} are the main conclusion of this work: the achievable delocalisation of the nanoparticle in this setup seems to be of the order of a few nanometers. This is smaller than the object's size, but much larger than the width of the ground state wavefunction ($z_{0,\text{NP}}\approx7~$pm, also plotted as a horizontal line in \cref{fig:res}). 
However, we cannot achieve complete ground state cooling for the nanoparticle (yet), this is considered in the next section. 

Another set of experimental parameters can provide a slight improvement in the induced superposition. For example, considering a nanoparticle charged with $300e$ instead of $800e$, results in a reduced separation between the nanoparticle and ion (for $V_\text{off} = 400\,\si{\volt}$ the separation is $43\,\si{\micro\metre}$ and $33\,\si{\micro\metre}$, for $Q\np = 800e$ and $Q\np=300e$, respectively).
This in turn results in a stronger Coulomb coupling (even though there are fewer charges on the nanoparticle) and the induced displacement of the nanoparticle from the Coulomb interaction overcomes the confining trapping potential more easily, allowing a larger displacement.
For $Q\np = 300e$ and the other experimental parameters unchanged, the stability conditions for the trapping (appendix~\cref{app:stability}) are still satisfied, but reducing the number of charges further would cause unstable trapping.
Changing the number of charges is more efficient for improving the displacement then changing the end-cap voltage, and it has the additional benefit that the coherence loss from charge-based interactions with the environment is reduced.

By manipulating the electric field potentials, a larger difference in the electric field strengths at the two superposition instances could be introduced to escalate the superposition size further. This has for example been done using magnetic fields in refs.~\cite{romero2017coherent,zhou2025spin}.
Alternatively, introducing a free-fall phase where the state evolves freely, could potentially increase the superposition size, as in Ref.~\cite{bateman2014near,PhysRevLett.107.020405,steiner2025free}; this has also been suggested as a measurement procedure~\cite{bateman2014near,toros_creating_2021}.

\subsection{Step 4: measurement scheme}\label{sec:measurement}

As described before, the measurement sequence we consider starts by by displacing the nanoparticle with an external uniform electric field, see~\cref{app:cphase}. Again neglecting the small squeezing originating from the Bogoliubov transformation, this results in \cref{eq:psi6a}. The phase depends on the nanoparticle position via $c_a^\uparrow$ and $c_a^\downarrow$. An order-1 control phase $\phi$ can be induced by applying a short-duration uniform electric field along the axial direction. 
Such a linear potential displaces the nanoparticle in phase space while picking up a relative phase proportional to the area enclosed in phase space (see~\cref{app:cphase}).
Controlling the strength of the uniform field $E_0$ and the time $t_E$ it is turned on, controls the control phase: $\phi<2\pi$ requires $E_0 t_E \sim 10^{-8} \si{\volt\second\per\metre}$.
Assuming a pulse duration of $10~\mu$s~\cite{Coherent-Bowler-2012} (i.e. $t_E\ll2\pi/\omega\np$), the voltage in the electrodes would be $\sim$mV. The corresponding effect on the ion is about $10^3$ smaller than that on the nanoparticle, and is thus neglected.

Finally, considering the `inverse SDK' on the ion motional state, we note that due to the additional displacement $c_b^{\uparrow,\downarrow}$ this SDK is not exactly of magnitude $\beta$ and may not recombine exactly both components to the ground state.

 \section{Beyond the ideal case}\label{sec:noise}
After having proved that a sizeable delocation of the wavefunction can be created on a relatively short timescale in the ideal case, we need to address some of the most important imperfections (notably relaxing the assumption that the nanoparticle is cooled to the ground state), as well as the unavoidable presence of decoherence.

\subsection{Effect on superposition size}
\begin{figure}[b]
    \centering
    \includegraphics[width=\linewidth]{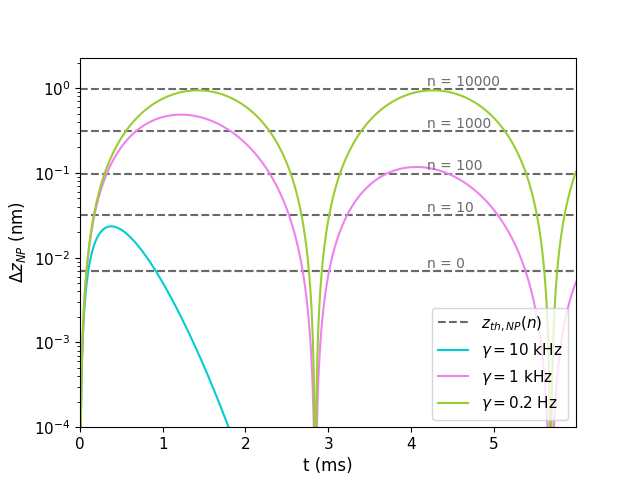}
    \caption{The induced superposition size in the nanoparticle as a function of time including amplitude damping. The experimental parameters are the same as in \cref{fig:res}, $\gamma=0.2$~Hz is taken from experimental studies in ref.~\cite{dania_ultra-high_2024}, and $\gamma=1,10$~kHz are plotted to illustrate the effect of damping on the superposition size. This figure considers an ion superposition of $100$~nm and would thus correspond to the green line in \cref{fig:res}.}
    \label{fig:res-improved}
\end{figure}
To account for the effects of damping on the nanoparticle due air molecule collisions,~\cref{fig:res-improved} shows it's oscillation with damping rate $\gamma = 0.2$~Hz. This is extracted from ref.~\cite{dania_ultra-high_2024}, which is measured at a pressure of $P\sim 10^{-4}$~mbar.
Ref.~\cite{dania_ultra-high_2024} showed the realization of a ultra-high quality nano-mechanical oscillator in a linear Paul trap, these measurements may not translate one-to-one to our setup, but give an impression of expected damping in a linear Paul trap. Additionally, \cref{fig:res-improved} includes damping rates at $\gamma=10^3, 10^4$~Hz damping rate to illustrate the effect of damping as a decrease in the maximum superposition size.

The stochastic displacements which will play a role in the decrease of the visibility in the next subsection, do not influence the superposition size in our model, since the noise is assumed to displace the two superposition arms equally and thus cancels out in the difference between the arms.

Previously, we had compared the superposition size to the ground-state width of the wavefunction. However, cooling the nanoparticle to the ground state is a very challenging endeavour. Cooling to the motional ground state has been achieved for trapped (calcium) ions in (linear) RF traps~\cite{Poulsen2012cooling,schmidt2000ground}, and for nanoparticles in optical traps~\cite{delic2020cooling,piotrowski2023simultaneous}, but the current state-of-the-art for nanoparticles in RF traps is cooling to $\sim$mK~\cite{Dania2021cooling}.
Therefore, \cref{fig:res-improved} also shows the width of the position uncertainty for thermal states given by $z_\text{th,NP} = z_{0,\text{NP}} \sqrt{1+2\bar{n}}$.
The figure shows that up to $\bar{n}=10^3$, the superposition size is significantly larger than that of the wavepacket width (for an initial ion superposition of $100$~nm and under a decoherence rate of $0.2$~Hz). This thermal occupation number would correspond to a temperature of the order $10~\mu$K.
At $\bar{n}=10^4$, the superposition size and width are approximately equal, this would correspond to a temperature of $T=0.1$~mK, which is a one order of magnitude improvement than the current state-of-the-art. 
These occupation numbers are strict constraints compared to what was reached in ref.~\cite{dania_ultra-high_2024} (a motional temperature of $17$~K along the $z$-axis).
Thus, executing the protocol proposed here would require great improvements in cooling; there are ideas on how to improve cooling in this specific setup~\cite{gupta2025quantum}.
As shown in the next subsection, the thermal state does not influence the visibility our proposed measurement protocol.

\begin{figure}[b]
    \centering
    \includegraphics[width=\linewidth]{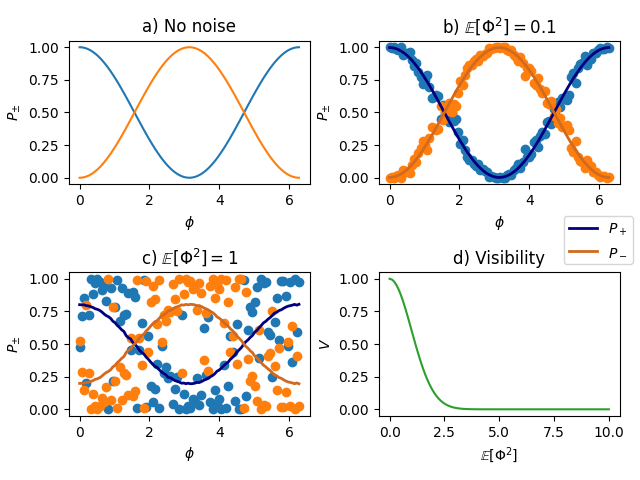}
    \caption{Probabilities of \cref{eq:P-noise}. (a) Without coherence loss. (b,c) Under coherence loss characterized by a stochastic phase following a Gaussian distribution with variance $\mathbb{E}[\Phi^2] = 0.1$ and $1$ respectively. Both the probabilities with random fluctuations (scatter plot) and the averaged noise (line plot) are shown.
    (d) Visibility of \cref{eq:P-noise} as a function of the variance, the plots b and c have $V=0.99$ and $0.60$, respectively. The physical interpretation of these chosen variances are discussed in the text and in~\cref{app:dephasing}.}
    \label{fig:visibility}
\end{figure}

\subsection{Effect on measurement}
Sources of coherence loss can reduce the visibility in the described measurement protocol, linear electric field noise as well as other noise sources (denoted by $\gamma$ below) may affect the probabilities in \cref{eq:prob} by
\begin{align}
    V= \frac{\text{max}P_+ - \text{min}P_+}{\text{max}P_+ + \text{min}P_+}=e^{-\frac{1}{2}\gamma}\label{eq:P-noise}   
\end{align}
For example, if the recombination of the ion state is not perfect, but leaves a difference $\delta \beta$, or if the timing is such that the nanoparticle motional states have a difference $\delta \alpha$, the visibility reduces with $\gamma=\abs{\delta \beta}^2, \abs{\delta \alpha}^2$, respectively (see~\cref{app:noise-meas}). 

This visibility is plotted as a function of the control phase in \cref{fig:visibility}a for $\gamma=0$.
If the nanoparticle is not prepared in the ground state but instead in a thermal state, we find that this has approximately no effect on the visibility because each instance in the mixture of coherent states evolves as the coherent state taken here.
To see this, suppose that the nanoparticle is initially prepared in a thermal state at temperature $T$:
\begin{equation}
    \rho_0(T) = \int_\alpha d^2\alpha \,p(\alpha,T) \ket{\alpha}\bra{\alpha}\np \otimes\ket{\psi_0}\bra{\psi_0}_\text{i}
\end{equation}
Where we have expressed the state of the nanoparticle as a mixture of coherent states in the Glauber–Sudarshan P representation. Here, $p(\alpha,T)$ denotes the thermal probability distribution in phase space at temperature $T$ and $\ket{\psi_0}_\text{i} = \left( \ket{\downarrow}_\text{i} \ket{-\beta}_\text{i} + \ket{\uparrow}_\text{i} \ket{\beta}_\text{i}\right) / \sqrt{2}\,$ is the prepared cat state of the ion. From the creation of the nanoparticle superposition to it's merger, the system undergoes the evolution:
\begin{align}
    \hat{U} &= \hat{U}_\text{i-SDK}\hat{U}_\text{control}\hat{U}_\text{C}=\mathbb{I}\np\otimes u_i
\end{align}
with:
\begin{align}
    u&=\exp{-i\phi(c_a^+,c_a^-)/2}\ket{\downarrow}\bra{\downarrow}\\
    &\quad + \exp{i\phi(c_a^+,c_a^-)/2}\ket{\uparrow}\bra{\uparrow}
\end{align} acting only on the spin of the ion. The phases $\phi(c_a^\uparrow,c_a^\downarrow)$ being the control phase from~\cref{app:cphase}. Therefore, the final state of the system is:

\begin{equation}
    \rho_f(T) =\int_\alpha d^2\alpha \,p(\alpha,T) \ket{\alpha}\bra{\alpha}\np \otimes (u\ket{\psi_0}\bra{\psi_0}u^\dagger)_\text{i}\,.
\end{equation}

We see that the final state of the ion upon readout has no dependence on the thermal distribution of the nanoparticle. Therefore, the coherence of the nanoparticle superposition can be observed even when it is not cooled to the ground state.
Here we have assumed that each state in the thermal mixture evolves identically. This is not exactly true since they differ in separation to the ion, however, it is a fair approximation as long as the spread of the the thermal state is much smaller than the separation between the ion and nanoparticle. This effect, that the thermal state does not influence a phase measurement, is also seen in a similar (Stern-Gerlach) interferometers \cite{wan_free_2016,dbrs-wn92}.

\Cref{fig:visibility} shows the probabilities $P_\pm$ as a function of the control phase $\phi$ for the noiseless and noise case, and the visibility as a function of noise.
The noise sources here are assumed to be stochastic and following a Gaussian noise distribution with zero mean.
We consider a Gaussian noise with a variance given by $\mathbb{E}[\Phi^2]$.
Considering linear force noise from electric fields, this noise can be seen as adding a perturbation to the control phase, i.e. $\phi+\Phi$, which, after averaging the noise over repeated experiments, gives the same exponential decay as in \cref{eq:P-noise} with $\exp(-\mathbb{E}[\Phi^2]/2)$ (i.e. $\gamma=\mathbb{E}[\Phi^2]$ in \cref{eq:P-noise}).
\Cref{app:dephasing} shows how the value $\mathbb{E}[\Phi^2]$ relates to the power spectral density for electric field noise and how it enters the probabilities. 
Figures (b) and (c) both plot the phase noise fluctuations as taken from a random distribution with variance $\mathbb{E}[\Phi^2]$ (the scatter plot); the solid line shows the averaged noise resulting in the exponential decay of the coherence terms as shown in \cref{eq:P-noise}.
This type of noise that leads to an exponential decay of the coherence term can also be sourced by voltage fluctuations, surface emitted electrons, and patch potentials that gives rise to linear electric field noise, as well as imperfect closure of the the interferometer arms. Vibrations of the trap setup can also cause acceleration noise that leads to dephasing.

\begin{table*}[ht]
\centering
\setlength{\tabcolsep}{20pt}
\def\arraystretch{1.5}
  \begin{tabular}{|l | l l | l l|}
  \hline
 \multirow{2}{*}{Particle ($\omega$)} & \multicolumn{2}{c|}{E-field noise ($\si{\volt\metre\per\sqrt\hertz}$)} & \multicolumn{2}{c|}{Acceleration noise ($\si{\metre\per\second^{2}\sqrt\hertz}$)} \\
  & threshold & literature & threshold & literature \\ [0.5ex]
 \hline
 NP ($1\,\si{\kilo\hertz}$) & $6\times10^{-8}$ & $4.1\times10^{-5}$~\cite{dania_ultra-high_2024} & $1\times10^{-8}$ & $5\times10^{-5}$~\cite{dania_ultra-high_2024} \\ 
 \hline 
 ion ($1\,\si{\mega\hertz}$) & $1\times10^{-7}$ & $3\times10^{-8}$~\cite{leopold2019cryogenic} & N/A & N/A \\ 
\hline
  \end{tabular}
  \caption{Electric field ($\sqrt{S_E}$) and acceleration ($\sqrt{S_a}$) PSD thresholds to keep decoherence under $0.1\,\si{\kilo\hertz}$. Based on \cref{eq:ion-psd,eq:np-psd}. Compared to experimental values in literature, note that ref.~\cite{leopold2019cryogenic} considers a $^9\text{Be}^+$ ion in a linear Paul trap with approximately the same secular frequency. 
  Ref.~\cite{dania_ultra-high_2024} considers a silica NP in a linear Paul trap, attributing all noise to either the electric field or acceleration noise. Literature values have only been given as an illustration and do not translate one-to-one to the current setup due to slight differences in e.g. the trap specifics. The ion acceleration noise is left out because at high frequencies we expect electric field noise to dominate.}
  \label{table:noise thresholds} 
\end{table*}

In~\cref{fig:visibility}, (a) plots the probabilities without noise, figure (b) considers $\mathbb{E}[\Phi^2]=0.1$, and figure (c) considers $\mathbb{E}[\Phi^2]=1$.
To put these values into context, if we consider the nanoparticle noise measured in ref.~\cite{dania_ultra-high_2024} to be sourced by electric field fluctuations ($\sqrt{S_E}\approx4.1\times10^{-5}\,\si{\volt\metre\per\sqrt\hertz}$) and e.g. the ion noise given by ref.~\cite{leopold2019cryogenic} ($\sqrt{S_{E}} < 3 \times 10^{-8}\,\si{\volt\metre\per\sqrt\hertz}$), these would combine to give $\mathbb{E}[\Phi^2]=1\times 10^{5}$ (see~\cref{app:dephasing}), which, from \cref{fig:visibility}d, can be seen to result is zero visibility.
This noise is dominated by the nanoparticle value from ref.~\cite{dania_ultra-high_2024}; this nanoparticle PSD value is three orders of magnitude larger than our constraint, indicating the need for improvement in order to carry out this protocol.

\Cref{table:noise thresholds} shows the acceptable electric field and acceleration noise required to keep the decoherence rate below $0.1\,\si{\kilo\hertz}$ at the respective oscillation frequencies of the nanoparticle and ion; the $0.1\,\si{\kilo\hertz}$ would correspond to $\mathbb{E}[\Phi^2]\approx0.3$, which has a visibility $V=0.95$. To compare these thresholds with currently achieved experimental values, we consider ref.~\cite{dania_ultra-high_2024}, which shows a Q-factor $>10^{10}$ for a levitated silica nanoparticle in a linear Paul trap at room temperature; they attributed the measured noise predominantly either to electric field noise or to acceleration noise arising from trap vibrations.
If the measured noise is dominated by electric field noise, the reported value is $\sqrt{S_E} = 4.1\times10^{-5}\,\si{\volt\metre\per\sqrt\hertz}$. If instead, the measured noise is dominated by vibrations, the reported acceleration noise is $\sqrt{S_a} = 5\times10^{-5}\,\si{\metre\per\second^{2}\sqrt\hertz}$. These values indicate that significant improvements in electric field shielding and vibration isolation at frequencies around $1\,\si{\kilo\hertz}$ are still required.
For the ion, the electric field noise measured in the trap of ref.~\cite{PhysRevLett.101.180602} at a frequency of around $1\,\si{\mega\hertz}$ corresponds to $\sqrt{S_{E}}\sim 1\times 10^{-7}\,\si{\volt\metre\per\sqrt\hertz}$. This suggests that current ion-trap technologies are already capable of achieving the required level of electric field noise suppression for the ion.

Additionally, the nanoparticle and ion interferometers need to be closed sufficiently well. For example, if the shift in position of the ion due to its interaction for the nanoparticle is not accounted for, this would result in dephasing given by $\abs{\delta \beta} = c_b^\uparrow-c_b^\downarrow \approx 44$, see \cref{eq:P-noise}. This would give a dephasing $\gamma\sim 10^3$, which will destroy all visibility, indicating that the 'inverse SDK' should be of different magnitude then the initial SDK to make sure the ion interferometer is closed sufficiently well, namely with $\abs{\delta \beta} < 0.3$ such that the dephasing is $<0.1$. Similarly, one can consider the timing precision of the `inverse SDK' such that the nanoparticle trajectory is closed with $\abs{\delta\alpha} < 0.3$, see \cref{eq:P-noise}. For the nanoparticle kHz oscillations we can assume the SDK to be instantaneous; the required timing precision is then within $0.04$~ms of the oscillation period. See~\cref{app:noise-meas}.

Spatial decoherence such as from blackbody photons and air molecules has been studied extensively~\cite{joos2013decoherence,PhysRevA.68.012105,schlosshauer2007decoherence,chang2010cavity}.
Assuming the geometric cross section, the decoherence rate from scattering with air molecules for an environmental temperature of $300\,\si{\kelvin}$ is $<0.1\,\si{\kilo\hertz}$ for pressures $<10^{-10}$~mbar (using the eqs. in~\cref{app:decoherence}).
Such a decoherence rate would conserve the coherence for at least one nanoparticle oscillation period.
The decoherence rate for blackbody scattering and absorption for the environmental temperature $300\,\si{\kelvin}$ is relatively small compared to the air molecule scattering, for nanometre-size superpositions: approximately $\sim 0.02$~kHz using the eqs. in~\cref{app:decoherence}.
Assuming that the nanoparticle is hotter than the environmental temperature due to the nanoparticle being loaded after the UHV has been achieved and the feedback cooling is done optically~\cite{bykov2024nanoparticle}, the blackbody emission decoherence rate $\sim0.1\,\si{\kilo\hertz}$ requires internal temperatures $<400\,\si{\kelvin}$ (again, see~\cref{app:decoherence}).
A quick calculation of the nanoparticle internal heating from the optical measurements used for the motional cooling shows an internal heating of $500$~K (assuming no heat loss, see~\cref{app:heating}), showing that blackbody emission may become troublesome and should be considered further.

The effect from multipoles in the material~\cite{PhysRevLett.133.253602}, emission of surface adsorbates~\cite{Stickler_desorption}, and a noisy electric field~\cite{PhysRevLett.134.033601}, on the rotational dynamics of electrically trapped particles~\cite{rademacher2025roto} has been studied both experimentally and theoretically; translation decoherence from dielectric response has also been studied~\cite{PRXQuantum.3.030327,PhysRevLett.83.4713}.

\section{Conclusion \& Discussion}
We have studied the possibility of engineering a delocalised state of a charged particle of the size of hundreds of nanometers (``nanoparticle'') through the interaction with an ion, based on the co-trapping in a Paul trap demonstrated in ref.~\cite{bykov2024nanoparticle}. We found that the coherent displacement of the nanoparticle can be of the orders of a few nanometers, smaller than the size of the nanoparticle itself but much larger than the spread of its initial ground state wavefunction.

The recent development co-trapping of an ion and nanoparticle in combination with the possibility of cooling~\cite{bykov20233d} made this setup attractive for exploring the creation of delocalised nanoparticle states. 
Having established that a significant delocalisation is in principle achievable with reasonable parameters, the next step is a comprehensive analysis of the sources of decoherence. Coherence may be lost through interaction with the environment (blackbody radiation, scattering with air molecules, Coulomb interaction with remnant environmental charges...) or from unavoidable fluctuations of the controls (static and oscillating fields of the trap, geometric deviations from the ideal...).
In~\cref{sec:noise} we have discussed several source of coherence loss and imposed constraints on the electric field noise.
As it stands, the current bottleneck is the noise isolation. The difference between our threshold and experimental values in literature based on similar trap designs shows that the shielding against noise needs to be improved by three orders of magnitude. This will be challenging and we hope this work serves as a motivation to work towards these benchmark values. For a full superposition measurements, cooling to near the motional ground will also become important, which has also not yet been realised in this platform.

Another relevant noise sources to this specific platform would be the micromotion and other RF specific noise sources. We leave this calculation for future work but note here that the axial trapping direction (along which the superposition is created) does not have dynamical trapping. Dynamical noise sources such as micromotion therefore only enter via either (Coulomb) coupling between the radial and axial directions, or via trapping imperfections, e.g. if the RF fields leak into the axial direction, or if the particles are not initialized at the radial centre of the trap. Such leakage has been seen in similar ion trapping experiments~\cite{leopold2019cryogenic}, and requires further investigation.

If this superposition were to be realized, it should be able to generate a measurable entanglement phase between two adjacent copies of this experiment via the Coulomb interaction, which would be the Coulomb equivalent of the QGEM experiment~\cite{bose2017spin} for testing gravitationally-induced entanglement.
Since this setup is a charged system, using it for small force measurements such as from the gravitational interaction or dark matter would be challenging due to charge-based coherence loss sources.
In terms of mass and superposition size, this proposal considers a higher mass and longer timescales compared to the proposal in a Talbot-Lau interferometers such as in ref.~\cite{nimmrichter_electron-enabled_2025} which uses electron-induced diffraction of a nanoparticle. Diamagnetic platforms often aim for larger superposition sizes, proposing methods for increasing an initial splitting of nanometre-size to micrometres, see e.g. ref.~\cite{zhou2025spin}, this platform currently sit at $\sim 2~\mu$m for a Bose-Einstein condensate of $10^4$ atoms 
within $0.5$~ms, and scaling up the mass remains challenging (as in any platform).

This proposal uses well-developed trapped-ion cat-states techniques to manipulate a levitated charged nanoparticle, thus avoiding direct control of the nanoparticle (internal) state. The platform is based on a recent experimental dual frequency trap which allows for co-trapping of the ion and nanoparticle. 
Although based on currently accessible techniques, this work shows that executing this proposal would require substantial improvements of noise isolation and cooling compared to the current state-of-the-art for this platform.

\section*{Acknowledgements}
This project is supported by the National Research Foundation, Singapore through the National Quantum Office, hosted in A*STAR, under its Centre for Quantum Technologies Funding Initiative (S24Q2d0009); and by the Ministry of Education, Singapore under the Academic Research Fund Tier 1 (FY2022, A-8000988-00-00).
\bibliography{bib} 
\bibliographystyle{IEEEtran}

\appendix

\section{For \cref{sec:setup}}

\subsection{Trapping stability}\label{app:stability}

Neglecting interactions between the ion and nanoparticle, the equations of motion for the dual Paul trap (see \cref{eq:eom-x-gen,eq:eom-z-gen}) can be written as Hill differential equations.
We briefly discuss the local stability of this setup.

As an example, we give the equations of motion in the x-direction, where, for simplicity, we assume that the frequency of the fast field is an integer multiple of the slow field: $\Omega_f = n \Omega_s$, $n\in\mathbb{N}$.
The substitution $T=\Omega_s t/2$ gives:
\begin{align}
    \ddot{x} + [a_x - 2 q_x \cos(2T) - 2 p_x \cos(2Tn)] x = 0 \, , \label{eq:x-eom-mat}
\end{align}
where we have defined the stability parameters as:
\begin{align}
    a_x &= - \left[ \frac{V_\text{off}}{r_0^2} + \frac{V_\text{end}}{z_0^2} \right] \frac{4 Q}{m \Omega_s^2}\,,\label{eq:ax}\\
    q_x &= \frac{2QV_s}{r_0^2 m \Omega_s^2}\, , \qq{} p_x = \frac{2QV_f}{r_0^2 m \Omega_s^2}\, , \label{eq:pq}
\end{align}
The derivative in \cref{eq:x-eom-mat} is now with respect to $T$.
Similarly, the equations of motion can be given as in \cref{eq:x-eom-mat} for the $y$ direction with $q_y = - q_x$, $p_y = - p_x$ and a relative minus sign in front of $V_\text{off}$ in $a_y$ compared to $a_x$.

The solutions to the Hill differential equations for the $x,y$ directions can be found using Floquet theory~\cite{abramowitzstegun}.
The stability of these solutions depends on the triplet $(a_i, p_i, q_i)$ with $i=x,y$.
The stability regime for the triplets such that the motion of the particle(s) is bounded can be found numerically.
Although not necessitated by these stability conditions, typical experimental realisations of linear Paul traps have small $a_i$.

Neglecting the fast voltage for the nanoparticle, the equations of motion become Matthieu-equations (i.e. Hill equations with a single RF frequency), which has well-understood solutions~\cite{abramowitzstegun,mclachlan1947theory}.
In the regime $\abs{q_i}\ll1$ where the pseudopotential approximation~\cite{dehmelt1968radiofrequency} is valid, one can approximate the trajectory of the trapped particle and find the frequency of the secular motion: $\omega_i = \sqrt{a_i+q_i^2/2}\Omega_s/2$.
In the chosen regime ($\abs{a_i}$, $\abs{q_i}\ll 1$), there is a large difference between the the frequencies $\omega_i \ll \Omega_s$; there is the secular motion frequency $\omega_i$ and the quick, small amplitude RF frequency called `micro-motion'~\cite{berkeland1998minimization} which acts as a perturbation to the secular motion.
It turns out that the micro-motion can often be neglected and that the bounded secular motion is harmonic in the pseudopotential approximation~\cite{Blatt:2003}.
In the case of the dual linear Paul trap, a similar psuedopotential approximation can also be used and two particles of different charge-to-mass ratios can be stably co-confined in the harmonic trap~\cite{Trypogeorgos2016,Leefer:2016zjr,bykov2024nanoparticle}

Here, we consider a nanoparticle ($R\np \sim 350\,\si{\nano\metre}$ and $Q\np=800e$) co-trapped with an atomic ion, such as realised in Ref.~\cite{bykov2024nanoparticle}.
The slow voltage dominates the trapping of the nanoparticle: taking only the slow voltage, the trapping frequency $\omega^\text{NP}_{x,y}$ is much larger than the trapping frequency when considering only the fast voltage~\cite{bykov2024nanoparticle}.
The frequency of the secular motion in the harmonic oscillator is therefore approximately:~\cite{abramowitzstegun,mclachlan1947theory,Blatt:2003}
\begin{equation}
    \omega^\text{NP}_{x,y} \approx \frac{1}{2}\Omega_s\sqrt{a_{x,y}^\text{NP} + (q_{x,y}^\text{NP})^2/2}
\end{equation}
where $a_{x,y}$ and $q_{x,y}$ are as given in \cref{eq:ax},~\cref{eq:pq} with $m=m\np$, $Q=Q\np$.
The stability condition is given by $0\leq a_{x,y} + q_{x,y}^2/2 \leq1$~\cite{Blatt:2003}, which in the regime that we are considering ($\abs{a_i}$, $\abs{q_i}\ll 1$) gives approximately: $\abs{a_{x,y}}\leq q_{x,y}^2/2$.
In the absence of offset and end-cap voltages this simply requires a non-zero slow field voltage.

For the ion, the charge-to-mass ratio is different: it is more sensitive to changes in the electric field and thus influenced more by the fast voltage, which has a smaller secular motion frequency, and, as a result, we cannot neglect the fast field. 
Instead, the slow potential is approximated as a static force (assuming $n$ is large, i.e. the slow and fast voltages are well-separated), such that the Hill equation becomes a Matthieu equation with modified $a_i$. 
For example, in the x-direction the equation of motion is given by (using $\tau = \Omega_f t / 2$ and denoting $\ddot{x}$ as the second derivative with respect to $\tau$):
\begin{align}
    &\ddot{x} + [a_x^\text{i} - 2 p_x^\text{i} \cos(2\tau)] x = 0 \\
    &a_x^\text{i} = - \left[ \frac{V_\text{off}}{r_0^2} + \frac{V_\text{end}}{z_0^2} \right] \frac{4 Q_\text{i}}{m_\text{i} \Omega_f^2} - \frac{4Q_\text{i}V_s}{r_0^2 m_\text{i} \Omega_f^2}\,, \label{eq:ion-ax}\\
    &p_x^\text{i} = \frac{2Q_\text{i}V_f}{r_0^2 m_\text{i} \Omega_f^2}\,  \label{eq:ion-px}
\end{align}
where the script i indicates the ion. 
Again, the Matthieu equation can be solved and the frequency of the harmonic oscillation is:~\cite{abramowitzstegun,mclachlan1947theory,Blatt:2003}
\begin{equation}
    \omega^\text{i}_{x,y} \approx \frac{1}{2} \Omega_f \sqrt{a_{x,y}^\text{i} + (p_{x,y}^\text{i})^2/2} \, .
\end{equation}
The stability conditions are given by $\abs{a_{x,y}^\text{i}}\leq (p_{x,y}^\text{i})^2/2$, as defined in \cref{eq:ion-ax,eq:ion-px}, $a_y^\text{i}$ has a relative minus sign for the offset and slow voltages compared to $a_x$ and $p_y = - p_x$.
\newline

For the $z$ direction, there is no AC voltage and the differential equation (with respect to $t$) is simple:
\begin{align}
    \ddot{z} + a_z z - g = 0 \,, \qq{} a_z = \frac{V_\text{end}}{z_0^2} \frac{2 Q}{m} \, .
\end{align}
Here, we have left out interactions with the nanoparticle, which will couple the nanoparticle and ion equations of motion.

Due to the RF fields being zero along the $z$-axis, the motion along $z$ is that of a harmonic oscillator with frequency
\begin{equation}
    \omega_z = \sqrt{a_z} \, ,
\end{equation}
which means that the axial field will always provide confinement as long as $a_z>0$ (and consequently $a_x + a_y < 0$).

The setup described here has been realised experimentally; see  ref.~\cite{bykov2024nanoparticle}.
To summarise, the motion of the particles is generally harmonic with micro-motions in the radial plane, and the stability conditions of the trapping depend on $Q/m$, $r_0$, $\Omega_{s,f}$, $V_{s,f}$ and $V_\text{end}$.
Following ref.~\cite{bykov2024nanoparticle}, these parameters are chosen such that the trapping is stable in the radial direction, as will be specified further in~\cref{sec:setup}.

\subsection{Ion-nanoparticle interactions}\label{app:interactions}
The previous discussion on trapping stability is performed in the absence of mutual interaction.
As mentioned in~\cref{sec:setup}, there is a strong Coulomb coupling between the ion and the nanoparticle.
Experimentally it was found that this interaction can be neglected for separations $>50\,\si{\micro\metre}$~\cite{bykov2024nanoparticle}.
In \cref{eq:lin-coulomb} we showed the linearized Coulomb interaction, which was used to find the induced displacement of the nanoparticle motional state, which happens at separations $<50\si{\micro\metre}$.
However, the Coulomb interaction is not the only electromagnetic interaction between the co-trapped particles, there are several other interactions, these are summarised in~\cref{table:int-strength} with their approximate relative strengths. 

Taking into account the dominant Coulomb interaction, in \cref{table:params} we found the centre of mass positions of the co-trapped particles numerically. 
Our numerical results match the experimental data shown in ref.~\cite{bykov2024nanoparticle}.

\begin{table}[ht]
\centering
\def\arraystretch{1.5}
\setlength{\tabcolsep}{7pt}
  \begin{tabular}{l c c}
 Interaction & relative strength & scaling \\ [0.5ex]
 \hline 
 \hline
 Coulomb & $1$ & $1/d$\\ 
 \hline 
 Dipole-Charge  & $10^{-6}$ & $1/d^2$\\
 \hline 
 Casimir  & $10^{-21}$ & $1/d^7$\\ 
 \hline 
 Magnetic dipole  & $10^{-15}$ & $1/d^3$ \\ 
\hline
  \end{tabular}
  \caption{Given the particle species with charges and masses given in \cref{table:params}. The separation is taken to be $20\,\si{\micro\metre}$, which is closer than the expected separation during our protocol and thus is a safe overestimation of the non-Coulomb forces, which scale more strongly with inverse separation ($d$). The dipole-charge is assumed to be between the trap-induced electric dipole in the nanoparticle and the ion charge, the Casimir interaction is taken between two spheres~\cite{casimir_influence_1948}, considering the atomic ion to be a polarisable medium.}
  \label{table:int-strength}
\end{table}

\section{For \cref{sec:quantum}}\label{app:supp}

\subsection{Quantization of the Coulomb interaction}\label{app:coulomb}

In~\cref{sec:displacement} we discuss the evolution of a quantum state under the Hamiltonian given by the self-interaction
\begin{align}
    &\hat{H_0}(d_\text{eq}) = \left[ Q\np \frac{V_\text{end}}{z_0^2} + \underbrace{\frac{1}{4\pi\epsilon_0}\frac{Q_\text{i} Q\np}{d_\text{eq}^3}}_\text{higher order} \right] \zn^2 \left( \a + \ad \right)^2 \nonumber \\
    &+ \left[ Q_\text{i} \frac{V_\text{end}}{z_0^2} + \underbrace{\frac{1}{4\pi\epsilon_0}\frac{Q_\text{i} Q\np}{d_\text{eq}^3}}_\text{higher order} \right] \zi^2 \left( \b + \bd \right)^2 \nonumber \\
    &+\left[ 2 Q\np \frac{V_\text{end}}{z_0^2} z^\text{eq}\np - \frac{1}{4\pi\epsilon_0}\frac{Q_\text{i} Q\np}{d_\text{eq}^2} + m g \right] \zn \left( \hat{a} + \hat{a}^\dagger \right) \nonumber \\
    &+ \left[ 2 Q_\text{i} \frac{V_\text{end}}{z_0^2} z^\text{eq}_\text{i}- \frac{1}{4\pi\epsilon_0}\frac{Q_\text{i} Q\np}{d_\text{eq}^2} + m g \right] \zi \left( \b + \bd \right) \nonumber \\
    & - \frac{\hbar\omega_{i}}{4} \left(\hat{b}^\dagger - \hat{b}\right)^2 - \frac{\hbar \omega\np}{4} \left(\hat{a}^\dagger - \hat{a}\right)^2\,, \label{eq:ham-ref1}
\end{align}
and of a bilinear interaction
\begin{align}
    \hat{H}_I^\text{CC}(d_\text{eq}) &= \underbrace{\frac{2 Q\np Q_\text{i}}{4\pi \epsilon_0 (d_\text{eq})^3} z_{0,i} z_{0,\text{NP}} \left(\hat{a}^\dagger + \hat{a} \right)\left(\hat{b}^\dagger + \hat{b} \right)}_\text{higher order} \,.\label{eq:ham-ref2}
\end{align} 
These Hamiltonians~\cref{eq:ham-ref1,eq:ham-ref2} depend on the equilibrium distance $d_\text{eq}$ which depends on the spin-state of the ion, this spin-dependence can be seen explicitly in \cref{eq:H-spin-dep}, where $\hat{H}(d_\text{eq}^{\uparrow,\downarrow})$ is given by $\hat{H}_0(d_\text{eq}^{\uparrow,\downarrow}) + \hat{H}_I(d_\text{eq}^{\uparrow,\downarrow})$.
We will neglect the indicated higher order terms in~\cref{eq:ham-ref1,eq:ham-ref2}, which are relatively small due to them being higher order terms of the expansion of the Coulomb interaction.
In this appendix we give the supporting calculations for the rewriting of the Hamiltonian in terms of generic coefficients.

\subsection{Vacuum shift}\label{app:calc-shift}
In \cref{eq:hd}, repeated above as \cref{eq:ham-ref1} plus \cref{eq:ham-ref2}, a vacuum shift is introduced, for a Hamiltonian of the shape
\begin{align}
    \hat{H} =&\, C_{12} \left(\hat{a}^\dagger + \hat{a} \right)\left(\hat{b}^\dagger + \hat{b} \right) + A_1 (\hat{a}^\dagger + \hat{a} ) + B_1 \left(\hat{b}^\dagger + \hat{b} \right) \nonumber \\
    &+A_2 (\hat{a}^\dagger + \hat{a})^2 + B_2 (\hat{b}^\dagger + \hat{b})^2  \nonumber \\
    & - M_2 \left(\hat{b}^\dagger - \hat{b}\right)^2 - M_1 \left(\hat{a}^\dagger - \hat{a}\right)^2     
\end{align}
the shifts $\hat{a}\to \hat{a} + c_a$ and $\hat{b}\to \hat{b} + c_b$ are defined as
\begin{align}
    c_a &= \frac{2 B_1 C_{12} - 4 A_1 B_2}{16 A_2 B_2 - (2C_{12})^2} \\
    c_b &= \frac{2 A_1 C_{12} - 4 A_2 B_1}{16 A_2 B_2 - (2C_{12})^2}
\end{align}
Note that the shifts are co-dependent.

In the case that $C_{12}=0$, the shifts are independent:
\begin{align}
    c_a^{\uparrow,\downarrow} = - \frac{A_1^{\uparrow,\downarrow}}{4 A_2}\,,\qq{} 
    c_b^{\uparrow,\downarrow} = - \frac{B_1^{\uparrow,\downarrow}}{4 B_2} 
\end{align}
However, the coefficients $A, B$ depend on the equilibrium distance $d_\text{eq}$ and thus the shifts $c_a$ and $c_b$ dependent on the spin-state of the ion.
In terms of the physical constants, neglecting smaller quadratic terms, $c_a$ is given by:
\begin{align}
    A_1^{\uparrow,\downarrow} &= \zn \bigg[ 2 Q\np \frac{V_\text{end}}{z_0^2} \azn - \frac{1}{4\pi\epsilon_0}\frac{Q_\text{i} Q\np}{(d_\text{eq}^{\uparrow,\downarrow})^2} \nonumber \\&\qq{}\qq{}+ m\np g \bigg]  \label{eq:A1}\\
    A_2 &= \zn^2 Q\np \frac{V_\text{end}}{z_0^2}  \label{eq:A2}
\end{align}
$c_b$ is similarly given for the ion. 
After this shift, only quadratic terms remain in the Hamiltonian and we perform a Bogoliubov transformation to simplify the expression further.

\subsection{Bogoliubov Transformation}\label{app:BT}
For a Hamiltonian
\begin{align*}
    \hat{H}_0 &= A_1 \left( \hat{a} \hat{a} + \hat{a}^\dagger \hat{a}^\dagger  \right) + A_2 \left(\hat{a}  \hat{a}^\dagger  + \hat{a}^\dagger  \hat{a}  \right) \\
    &\qq{}+ B_1 \left( \hat{b} \hat{b}  + \hat{b}^\dagger \hat{b}^\dagger  \right) + B_2 \left(\hat{b}  \hat{b}^\dagger  + \hat{b}^\dagger  \hat{b}  \right) + \tilde{C} \\
    \hat{H}_1 &= C_{12} \left(\hat{a}^\dagger  + a  \right)\left(\hat{b}^\dagger  + \b \right)
\end{align*}
The Bogoliubov transformation is given by:
\begin{align*}
    &\begin{pmatrix}
    \tilde{a} \\
    \tilde{a}^\dagger 
    \end{pmatrix}
    = 
    \begin{pmatrix}
    \cosh(r_a) & \sinh(r_a) \\
    \sinh(r_a) & \cosh(r_a)
    \end{pmatrix}
    \begin{pmatrix}
        \a \\
        \ad
    \end{pmatrix}
     \,, \\
    &\begin{pmatrix}
    \tilde{b} \\
    \tilde{b}^\dagger 
    \end{pmatrix}
    = 
    \begin{pmatrix}
    \cosh(r_b) & \sinh(r_b) \\
    \sinh(r_b) & \cosh(r_b)
    \end{pmatrix}
    \begin{pmatrix}
        \b \\
        \bd
    \end{pmatrix}
    \\
    &\text{with } 
    \tanh(2r_a) = \frac{A_1}{A_2} \,\,\text{ and } \,\, \tanh(2r_b) = \frac{B_1}{B_2} 
\end{align*}
The frequencies are found to be
\begin{equation}
    \tilde{\omega}_{a} = 2\sqrt{A_2^2 - A_1^2} \,,\qq{} \tilde{\omega}_{b} = 2\sqrt{B_2^2 - B_1^2}
\end{equation}
and the shifted coupling constant is:
\begin{align}
    \tilde{C}_{12} = C_{12} \sqrt{1 - \frac{A_1^2}{A_2^2}} \sqrt{1 - \frac{B_1^2}{B_2^2}} \left( 1 - \frac{A_1}{A_2}\right)\left(1-\frac{B_1}{B_2}\right)
\end{align}
Resulting in the transformed Hamiltonian:
\begin{equation}\label{eq:H-BT}
    \hat{H} =  \tilde{\omega}_a \hat{a}^\dagger\hat{a} + \tilde{\omega}_b \hat{b}^\dagger\hat{b} + \tilde{C}_{12} \left(\hat{a}^\dagger + \hat{a} \right)\left(\hat{b}^\dagger + \hat{b} \right) + \bar{C}
\end{equation}
Note, that, from the expression in terms of physical parameters it is the case that $A_1^2<A_2^2$ and thus $\tilde{C}_{12}\in\mathbb{R}$ and the Hamiltonian is Hermitian. 

Including the higher order terms in \cref{eq:ham-ref1,eq:ham-ref2} will cause the effective frequencies $\tilde{\omega}_{a,b}$ to become dependent on the separation:
\begin{align*}
    \tilde{\omega}_a^{\uparrow,\downarrow} &= 2 \zn \sqrt{\hbar Q\np \omega\np \left[ \frac{V_\text{end}}{z_0^2} + \frac{1}{4\pi\epsilon_0}\frac{Q_\text{i}}{(d^{\uparrow,\downarrow}_\text{eq})^3} \right] }\\
    \tilde{\omega}_b^{\uparrow,\downarrow} &= 2 \zi \sqrt{\hbar Q_\text{i} \omega_\text{i} \left[ \frac{V_\text{end}}{z_0^2} + \frac{1}{4\pi\epsilon_0}\frac{Q\np}{(d^{\uparrow,\downarrow}_\text{eq})^3} \right]}
\end{align*}
This shifts the frequency slightly, $\tilde{\omega}_a^\uparrow/\tilde{\omega}_a^\downarrow = 0.99998$, $\tilde{\omega}_b^\uparrow/\tilde{\omega}_b^\downarrow = 0.99511$. In~\cref{sec:displacement} we have done the derivation of the relative displacement assuming that $\tilde{\omega}_{a,b}^\uparrow \approx \tilde{\omega}_{a,b}^\downarrow$ and neglecting the coupling $\tilde{C}_{12}$ (since $\tilde{C}_{12}\ll\tilde{\omega}_{a,b}$).
Performing the derivation without these assumptions is not straightforward because the rotating-wave approximation ($\tilde{\omega}_a - \tilde{\omega}_b \ll \tilde{\omega}_a + \tilde{\omega}_b$) does not hold ($\tilde{\omega}_a/ \tilde{\omega}_b \sim 10^{-3}$), which is clear from the the expressions of $\tilde{\omega}_a$, $\tilde{\omega}_b$ above for the experimental parameters set out in \cref{sec:setup}.
Since the coupling is given by a higher-order Coulomb term, we leave the full derivation for future work and only consider the linearised Coulomb approximation that results in \cref{eq:delz-res}.

\section{For \cref{sec:noise}}\label{app:dephasing}
We discuss several source of noise to support the main text: (1) the dephasing due to electric field fluctuations, (2) internal heating due to motional cooling, and (3) spatial decoherence.

\subsection{Readout Control Phase}\label{app:cphase}
The control phase is generated by ensuring that the nanoparticle encloses an area in phase space throughout the protocol \cite{Chen_Ni_Shen_1993}. 
By applying a control electric field $E_c$ for a time of $t_c$ after the nanoparticle superposition is created, it undergoes a displacement in phase space:
\begin{equation}
\begin{split}
    \xi_c &= -i Q\np E_c\sqrt{\frac{1}{2 \hbar m\np \omega\np}} \int_{0}^{t_c} dt' e^{i\omega\np t'}\\
    &= - Q\np E_c\sqrt{\frac{1}{2 \hbar m\np \omega^2\np}} (e^{i\omega\np t}-1)
\end{split}
\end{equation}
After the nanoparticle's motional state is merged, it acquires the control phase proportional to the area $\Im{\xi_c \Delta c}$ that it encloses in phase space:
\begin{equation}
\begin{split}
    \phi(c_a^\uparrow,c_a^\downarrow) &= -2\Im{\xi_c \Delta c}\\
    &= - Q\np E_c \Delta c\sqrt{\frac{2}{ \hbar m\np \omega^2\np}} \sin(\omega\np t_c)
\end{split}
\end{equation}
Where $\Delta c = c_a^\uparrow - c_a^\downarrow$
For $t_c \ll \frac{2\pi}{\omega\np}$, we have:
\begin{equation}
    \phi(c_a^\uparrow,c_a^\downarrow) \approx - Q\np E_c \Delta c\sqrt{\frac{2}{ \hbar m\np \omega^2\np}} \omega\np t_c
\end{equation}

In order to acquire a phase of order $\sim 1$, we therefore require $E_0 t_E \sim 10^{-8} \si{\volt\second\per\metre}$.

\subsection{Readout under noise}\label{app:noise-meas}
As discussed above, the relative phase can cause dephasing, this can also show up in the measurement protocol. 
The relative phase $\Phi$, like the control phase (which we denote $\phi$ here but is understood to depend on the nanoparticle position; $\phi(c_a^\uparrow,c_a^\downarrow)$), will enter the probabilities in the coherence times: 
\begin{equation}
    P_\pm = \frac{1}{2} \pm \frac{1}{2} \cos(\phi+\Phi)
\end{equation}
Averaging over the noise realization gives:
\begin{align}
    P_\pm 
    &= \frac{1}{2} \pm \frac{1}{2} \mathbb{E}\left[ \frac{1}{2} e^{i (\phi+\Phi)} + \frac{1}{2} e^{-i (\phi+\Phi)} \right] \\
    &= \frac{1}{2} \pm \frac{1}{2} \left( \frac{1}{2} e^{i \phi} \mathbb{E}\left[e^{i\Phi}\right] +\frac{1}{2} e^{-i \phi} \mathbb{E}\left[e^{-i\Phi}\right]\right) \\
    &= \frac{1}{2} \pm \frac{1}{2} e^{-\mathbb{E}[\Phi^2]/2} \cos(\phi)
\end{align}
Showing an exponential decay of the coherence terms characterized by the variance of $\Phi$ (which is expressed in detail in the previous section).

Similarly, for a final state (after recombination): 
\begin{align}
    \ket{\Psi} \propto \big[ & \ket{0}\np\ket{0}_\text{ion}  \ket{\downarrow} + e^{i\phi} \ket{\delta \alpha}\np\ket{\delta \beta}_\text{ion}  \ket{\uparrow}\big]
\end{align}
where the $\delta \beta$ and $\delta \alpha$ are imperfections in the final state due to e.g. an imperfect reversal of the ion SDK or a timing error in recombining the superpositions.
The change to the $\{\ket{+},\ket{-}\}$ basis with $\ket{\pm} = (\ket{\downarrow}\pm\ket{\uparrow})/\sqrt{2}$ gives:
\begin{align}
    \ket{\Psi} \propto &\left( \ket{\delta \beta}\np\ket{\delta\alpha}_\text{ion}  + e^{i\phi} \ket{0}\np \ket{0}_\text{ion} \right) \ket{+} \nonumber \\ &- \left( \ket{\delta \beta}\np\ket{\delta\alpha}_\text{ion} - e^{i\phi} \ket{0}\np \ket{0}_\text{ion} \right) \ket{-}
\end{align}
As a result of the identity for the overlap of coherent states, the probabilities pick up an exponential decay in the coherence terms:
\begin{align}
    P_\pm 
    &=\frac{1}{2} \left( 1 \pm \bra{\delta \alpha}\ket{0} \bra{\delta \beta}\ket{0} \cos(\phi) \right) \\
    &=\frac{1}{2} \left( 1 \pm e^{-\frac{1}{2}\abs{\delta\alpha}^2} e^{-\frac{1}{2}\abs{\delta\beta}^2} \cos(\phi) \right)
\end{align}
showing the exponential decay of coherence.
The value $\delta\alpha$ could for example be caused by a timing of the inverse-SDK when the nanoparticle states do not overlap, it scales as $\cos(\omega t)$, following \cref{eq:delz-res}.

\subsection{Dephasing from Electric Field and Acceleration Noise}
We shall consider the effects of dephasing on the readout spin state due to electric field fluctuations and acceleration noise (caused by mechanical vibrations of the trap). These two noise sources would be modelled as stochastic uniform forces on the particles in a harmonic trap, from which we derive a transfer function to relate their power spectral densities (PSD) to the dephasing rate. This would then give us the acceptable amount of noise that the protocol can tolerate (to keep the dephasing rate under $\sim 0.1\text{ kHz}$, set by the oscillation frequency of the nanoparticle). 

Under uniform force noise, the nanoparticle/ion Hamiltonian along the trap axis during free evolution would be:
\begin{equation} \label{hamiltonian with uniform force noise}
    \hat{H}(t) = \hat{H}_0 +  \lambda(t) \hat{x}
\end{equation}
where $\lambda(t)$ is a stochastic force. For electric field noise $E(t)$, $\lambda(t)=QE(t)$ and for acceleration noise $a(t)$, $\lambda(t)=ma(t)$, where $Q$ and $m$ are respectively the charges and mass of the nanoparticle/ion. The time independent term is:
\begin{equation}
    \hat{H}_0 = \frac{\hat{P}^2}{2m} + \frac{1}{2}m\omega^2\hat{x}^2
\end{equation}
This is a quantum forced harmonic oscillator, for which the evolution operator in the interaction picture is given by \cite{dbrs-wn92}:
\begin{equation}
    \hat{U}_I(t_0,t) = e^{i\varphi(t_0,t)} \hat{D} (\xi(t_0,t))
\end{equation}
where $\xi(t)$ and $\varphi(t)$ can be solved by the integral equations:
\begin{equation}\label{displacement noise}
    \xi(t_0,t) = -i \sqrt{\frac{1}{2 \hbar m \omega}} \int_{t_0}^t dt' \lambda(t') e^{i\omega t'}
\end{equation}
\begin{equation}
    \varphi(t_0,t)  = \Im\left(\int_{t_0}^{t} d\xi^*(t') \xi(t')\right)
\end{equation}
the phase $\varphi(t)$ is twice the area enclosed by $\xi(t)$ in phase space.
The force noise leads to displacement noise on the particles.

Taking into account the force noise as characterized by a displacement $D(\xi_\text{i})$ on the ion and $D(\xi\np)$ on the nanoparticle, the final state before readout would therefore be:
\begin{equation}
\begin{split}
    \ket{\Psi_0'} &\propto \hat{D}_\text{i}(\xi_\text{i})
    \otimes\hat{D}\np(\xi_\text{NP})\ket{-c_a^-}\np\ket{-\beta - c_b^-}_\text{ion}\ket{\downarrow} \nonumber\\+ &e^{i\phi(c_a^+,c_a^-)} \hat{D}_\text{i} (\xi_\text{i})\otimes\hat{D}\np(\xi_\text{NP})\ket{-c_a^+}\np\ket{\beta - c_b^+}_\text{ion}\ket{\uparrow} 
\end{split}
\end{equation}
We take $\xi_\text{i} = \xi(t_i,t_f)$ and $\xi_\text{NP} = \xi(t_i,t_f)$, where $t_i$, and $t_f$ are the time of creation of the ion (NP) superposition and the merger of the ion (NP) superposition, respectively.
In the noise analysis the dynamics of the creation/annihilation of the cat states are not taken into account, thus providing an upper bound on the dephasing~\cite{Myatt2000-qp}.
The merger of the nanoparticle superposition from coulomb interaction with the ion superposition, followed by the application of the SDK to merge the ion superposition leads to:
\begin{align*}
    \ket{\Psi_1}&= \;\hat{U}_\text{i-SDK}\ket{\Psi_0'}\\ \propto &\ket{\xi\np}\np\ket{\xi_\text{i}}_\text{ion}\left(\ket{\downarrow}+e^{-i\phi(c_a^+,c_a^-)}e^{i\Phi}\ket{\uparrow}\right)
\end{align*}
where:
\begin{align}
    &\Phi = -2\Im(\xi\np\Delta c_a) -2\Im(\xi_i(2\beta - \Delta c_b)) \label{efield phase noise} \\
    &\Delta c_a \equiv c_a^+-c_a^-, \quad \Delta c_b \equiv c_b^+-c_b^-
\end{align}
The relative phase $\Phi$ comes from the non-commuting nature of the displacement operators:
\begin{equation}
    \hat{D}(\beta)\hat{D}(\xi) = e^{-i2\Im{\xi\beta^*}} \hat{D}(\xi)\hat{D}(\beta)
\end{equation}
As a relative complex phase, this does not cause a `coherence loss' necessarily, since the magnitude of the off-diagonal elements does not decay when multiplied by a complex phase.
However, over repeated experiments these fluctuations cause a decay in visibility which is shown as an exponential decay in the coherence terms when averaging the density matrix.
The averaged density matrix of the readout spin state (considering perfect recombination of the motional states) is:
\begin{equation}
\mathbb{E}[\rho_s] = 
\begin{bmatrix}
    1 & e^{i\phi(c_a^+,c_a^-)}\mathbb{E}[e^{-i\Phi}] \\
    e^{-i\phi(c_a^+,c_a^-)}\mathbb{E}[e^{i\Phi}] & 1 
\end{bmatrix}
\end{equation}
where the statistical average $\mathbb{E}[.]$ is taken over different instances of $\lambda(t)$. We see that the off-diagonal terms decay with $\mathbb{E}[e^{i\Phi}]$, where assuming a zero mean value of $\xi\np$ and $\xi_i$, gives $\mathbb{E}[e^{i\Phi}] = e^{-\mathbb{E}[\Phi^2]/2}$~\cite{wiener1930generalized,khintchine1934korrelationstheorie}. 
Hence, the off-diagonals decays exponentially with the variance of $\Phi$. Also assuming that $\xi\np$ and $\xi_i$ are not correlated, we then plug in eq.~\cref{displacement noise,efield phase noise} to get $\mathbb{E}[\Phi^2]$ in terms of $\lambda(t)$ (see \cref{hamiltonian with uniform force noise}):
\begin{equation}
\begin{split}
    \mathbb{E}[\Phi^2] &= \bigg(\mathbb{E}[4\Im(\xi\np\Delta c_a)^2]\\
    & \quad +  \mathbb{E}[4\Im(\xi_i(2\beta - \Delta c_b))^2]\bigg)\\
    &= \Gamma\np(\xi\np,\Delta c_a) + \Gamma_\text{i}(\xi_i, 2\beta - \Delta c_b)
\end{split}
\end{equation}
where we define:
\begin{equation}\label{eq:gamma1}
\begin{split}
    \Gamma_\mu(\xi,\alpha) &= \mathbb{E}[4\Im(\xi \alpha)^2]\\
    &= \frac{16\alpha^2}{2\hbar m_\mu\omega_\mu} \int_{t_i}^{t_f} \int_{t_i}^{t_f} dt_1 dt_2 \mathbb{E} [\lambda(t_1)\lambda(t_2)] \\ &\qq{}\qq{}\qq{}\,\,\,\,\,\,\,\times\cos(\omega_\mu t_1)\cos(\omega_\mu t_2)
\end{split}
\end{equation}
Here, $\mu= \text{i}, \text{NP}$ denotes the parameters to be from either the nanoparticle or ion.
By the Wiener-Khinchin theorem~\cite{chatfield1989analysis}, we can express the auto-correlation function of the force $\lambda(t)$ in terms of its PSD $S_\lambda(\Omega)$ (where $\Omega$ is the frequency of the force noise):
\begin{equation}\label{eqn: noise autocorrelation}
\begin{split}
    \mathbf{E} [\lambda(t_1)\lambda(t_2)] &= \int_0^\infty \frac{d\Omega}{2\pi} S_\lambda (\Omega) e^{i \Omega (t_1-t_2)}
\end{split}
\end{equation}

This allows us to express $\Gamma_\mu(\xi,\alpha)$ in terms of the force noise PSD and a transfer function $F_\mu(\Omega,\xi,\alpha) $, from \cref{eq:gamma1}:
\begin{equation}
    \Gamma_\mu(\xi,\alpha) = \int_0^\infty d\Omega  S_\lambda (\Omega) F_\mu(\Omega,\xi,\alpha) 
\end{equation}
with
\begin{align}
    &F_\mu(\Omega,\xi,\alpha) \nonumber \\ 
    &= \frac{4\alpha^2}{\pi\hbar m_\mu\omega_\mu} \int_{t_i}^{t_f} \int_{t_i}^{_f} dt_1 dt_2 e^{i \Omega (t_1-t_2)}\cos(\omega t_1)\cos(\omega t_2) \nonumber \\
    &= \frac{\alpha^2}{\pi\hbar m_\mu\omega_\mu} \bigg|\frac{e^{i(\Omega+\omega_\mu)t_f}-e^{i(\Omega+\omega_\mu)t_i}}{\Omega+\omega_\mu}\quad \nonumber \\
    & \qquad \qquad\qquad +\frac{e^{i(\Omega-\omega_\mu)t_f}-e^{i(\Omega-\omega_\mu)t_i}}{\Omega-\omega_\mu}\bigg|^2 \label{eq:Fmu}
\end{align}
We see that the noise at frequency $\Omega \approx\omega_\mu$ dominates the dephasing (as $\Omega\to\omega_\mu$, the second term in \cref{eq:Fmu} blows up).
Hence, we can approximate:
\begin{align}
        \Gamma_\mu(\xi,\alpha) &\approx S_\lambda (\omega_\mu)\int_0^\infty d\Omega   F_\mu(\Omega,\xi,\alpha) \nonumber\\
        & \approx S_\lambda (\omega_\mu)\frac{\alpha^2}{\pi\hbar m_\mu\omega_\mu}\int_0^\infty d\Omega\abs{\frac{e^{i(\Omega-\omega_\mu)(t_f-t_i)}-1}{\Omega-\omega_\mu}}^2 \nonumber \\
        &= S_\lambda (\omega_\mu)\frac{2\alpha^2}{\hbar m_\mu\omega_\mu} (t_f-t_i) \label{eq:gamma2}
\end{align}
With the assumption $\Omega\approx\omega_\mu$, i.e. selecting the dominant modes in the transfer function, we have made the substitution $S_\lambda(\Omega)\to S_\lambda(\omega_\mu)$, i.e. we consider the PSD only at the oscillation frequency.
Note that the frequency-scaling of the PSD depends on the noise source, e.g. an unshielded ion trap typically scales with $S_E(\Omega)\sim\Omega^{-3}$ due to electromagnetic interference~\cite{brownnutt_ion-trap_2015}.
Based on \cref{eq:gamma2}, the dephasing rate is given by:
\begin{equation}
    \frac{d\Gamma_\mu(\xi,\alpha)}{dt} =  S_\lambda (\omega_\mu)\frac{2\alpha^2}{\hbar m_\mu\omega_\mu}
\end{equation}
To maintain a dephasing $<0.1\text{ kHz}$, we would thus require the following constraints on the force noise PSD:

\begin{align}
    &S_\lambda (\omega_\text{i}) < \frac{\hbar m_\text{i}\omega_\text{i}}{2(2\beta - \Delta c_b)^2} 100 \label{eq:ion-psd} \\
    &S_\lambda (\omega\np) < \frac{\hbar m\np\omega\np}{2(\Delta c_a)^2} 100 \label{eq:np-psd}
\end{align}
This force noise can originate from e.g. electric field fluctuations or vibrational noise, giving a constraint on the electric field PSD via the relation $S_E (\Omega) = S_\lambda (\Omega)/Q^2$ and the acceleration noise PSD via $S_a (\Omega) = S_\lambda (\Omega)/m^2$. 
We can derive the PSD thresholds for the electric field noise and acceleration noise respectively, this gives the values shown in the main text. 
Finally, as a common source of electric field noise, the relation to voltage noise can be modelled as $S_V(\omega) = d\,S_E (\omega)$~\cite{brownnutt_ion-trap_2015}, where $d$ is the distance between the electrodes sourcing the noise.

\subsection{Internal heating from motional cooling}\label{app:heating}
The internal temperature of the nanoparticle is not equal to the environmental temperature inside the vacuum chamber. 
For example, optical detection is used for the cooling of the motional state of the nanoparticle, which causes heating.
The change in temperature ($\dd{T}$) is given by the transferred heat ($\dd{Q}$) is given by the volumetric heat capacity ($c_V$): 
\begin{equation}
    \dd{Q} = c_V V \dd{T}
\end{equation}
The change in transferred heat can be expressed as the absorbed power, which for a nanoparticle of uniform temperature illuminated by a monochromatic laser of intensity $I$ is:
\begin{equation}
    \dv{Q}{t} = P_\text{absorption} = I \sigma_\text{abs}
\end{equation}
where $\sigma_\text{abs}$ is the absorption cross section. 
The intensity of the light attenuates as it propagates through a medium: 
\begin{align}
    &I(d) = I(0) e^{-\alpha d} \,\Rightarrow\, 1-\frac{I(d)}{I(0)} \approx 1 - 1 + \alpha d \\&\Rightarrow\, \alpha 2 R \text{ is the fraction absorbed}
\end{align}
for small absorption approximation.
Here, $\alpha$ is the attenuation coefficient, related to the complex part of the reflective index, $\kappa$ (i.e. the loss in the reflection), the value here taken from ref.~\cite{bateman2014near}: 
\begin{equation}
    \alpha = \frac{4\pi \kappa(\lambda)}{\lambda}
\end{equation}
As an approximation we take the effective absorption cross section to be the geometrical cross section times the fraction that is absorbed:
\begin{equation}
    \sigma = \pi R^2 \alpha 2 R 
\end{equation}
and thus 
\begin{equation}\label{eq:pabs}
    P_\text{abs} = I \, 2 \pi R^3 \alpha \approx 10^{-19} \text{W}
\end{equation}
The power of the laser and the waist of the beam are given in ref.~\cite{dania_ultra-high_2024}: $21$~mW and $300~\mu$m respectively, which is at $780$~nm.
The intensity can then be found via $I = 2 P / (\pi w^2)$ (peak intensity for Gaussian beam).
Then:
\begin{align}
    \Delta T = \frac{P_\text{abs}}{c_V V} \approx 460 \text{K} \\\nonumber
\end{align}
For a ms of illumination. Of course, this is an approximation accounting only for absorption and not for the subsequent emission of photons due to the heating.
Additionally, any defects in the material are neglected in this approximation.

There are other sources of heating, such as photon absorption from heated surfaces, which can also contribute to an increase of the internal temperature of the nanoparticle.

\subsection{Spatial decoherence}~\label{app:decoherence}
To estimate the coherence times limited by blackbody scattering/emission/absorption and scattering with air molecules we consider the decoherence rate derived in e.g.~\cite{schlosshauer2007decoherence,romero2011quantum}, which takes the blackbody to be in the long-wavelength limit and air molecule scattering to be in the short-wavelength limit compared to the superposition size. 
The latter being valid for $T>$nK and the former for $T<10^6$~K (comparing the thermal de Broglie wavelength with $\sim1$~nm).
\begin{align}
    \gamma_\text{air} &= 2 n_V R^2 \sqrt{\frac{2 \pi T_e k_b}{m_\text{air}}} \label{eq:air} \\
    \Lambda_\text{bbs} &=  8! \zeta(9) \frac{8 R^6 c}{9\pi} \abs{\frac{\epsilon-1}{\epsilon+2}}^2 \left( \frac{k_b T_e}{\hbar c}\right)^9 \label{eq:bbs}\\
    \Lambda_\text{bbe(a)} &= \frac{16 \pi^5 c R^3}{189} \Im\left(\frac{\epsilon-1}{\epsilon+2}\right) \left( \frac{k_b T_{e(i)}}{\hbar c}\right)^6 \label{eq:bbae}
\end{align}
where the blackbody scattering (bbs) and blackbody absorption (bba) use the environment temperature and the blackbody emission (bbe) uses the internal temperature of the nanoparticle.
Here, $n_V$ is the number density of air molecules, $\zeta$ is the Riemann zeta function and $\epsilon_r$ is the relative permittivity.

\end{document}